\documentclass[prl,floatfix,reprint]{revtex4}
\usepackage{graphicx,amssymb,amsmath}

\begin{document}

\title{Anatomy and efficiency of urban multimodal mobility}

\author{Riccardo Gallotti}
\affiliation{Institut de Physique Th\'{e}orique, CEA, CNRS-URA 2306, F-91191, 
Gif-sur-Yvette, France}
\author{Marc Barthelemy\footnote{Correspondance to: marc.barthelemy@cea.fr}}
\affiliation{Institut de Physique Th\'{e}orique, CEA, CNRS-URA 2306, F-91191, 
Gif-sur-Yvette, France}

\begin{abstract} 

The growth of transportation networks and their increasing interconnections, although positive, has the downside effect of an increasing complexity which make them difficult to use, to assess, and limits their efficiency. On average in the UK, $23\%$ of travel time is lost in connections for trips with more than one mode, and the lack of synchronization decreases very slowly with population size. This lack of synchronization between modes induces differences between the theoretical quickest trip and the `time-respecting' path, which takes into account waiting times at interconnection nodes. We analyse here the statistics of these paths on the multilayer, temporal network of the entire, multimodal british public transportation system. We propose a statistical decomposition -- the `anatomy' -- of trips in urban areas, in terms of riding, waiting and walking times, and which shows how the temporal structure of trips varies with distance and allows us to compare different cities. Weaknesses in systems can be either insufficient transportation speed or service frequency, but the key parameter controlling their global efficiency is the total number of stop events per hour for all modes.  This analysis suggests the need for better optimization strategies, adapted to short, long unimodal or multimodal trips. 

\end{abstract}

\maketitle

\section{Introduction}

Although the coupling between different transportation networks is fundamental~\cite{Morris:2012}, most of the studies on Public Transport Networks have been performed considering only one single transportation mode: private cars~\cite{Bazzani:2010, Gallotti:2012, Gallotti:2013}, taxis~\cite{Bin:2009, Leung:2011, Liang:2012, Liu:2012}, Subway~\cite{Latora:2001, Angeloudis:2006, Lee:2008, Derrible:2010, Roth:2012}, Train~\cite{Sen:2003, Seaton:2004, Kurant:2006, Kurant:2006b, Li:2007, Drobritz:2009}, Bus and Trams~\cite{Sienkiewicz:2005, Kurant:2006b, Xu:2007, Chen:2007}, and at a worldwide scale, airline networks (see~\cite{Zanin:2013} and references therein). However, most transportation systems are coupled to each other and as it was recently shown in ~\cite{Buldyrev:2010}, interconnections can have dramatic consequences on the behavior of the whole system. This finding triggered a wealth of studies~\cite{Mucha:2010, DeDomenico:2013, Gomez:2013, Nicosia:2013, Kivela:2014, Boccaletti:2014} on multilayer networks --- also coined multiplex networks --- providing a new paradigm for studying these coupled systems. Public Transport Networks belong to this class and provides a paradigmatically example of spatial~\cite{Barthelemy:2011}, temporal~\cite{Holme:2012}, and multilayer Network~\cite{Kivela:2014} where each layer corresponds to a single transportation mode. 

A few studies only considered many modes merged in an unique network~\cite{vonFerber:2009}, but this aggregation might hide important structural features due to the intrinsical multilayer nature of the network~\cite{Cardillo:2013}. In particular, in the case of urban transport, not considering the connection times can lead to unprecise estimates for the network's navigability\cite{DeDomenico:2014}. We note also that interchanges are not symmetrical: rail-to-bus and bus-to-rail waiting time are different and are independent from the actual traffic volume~\cite{Coffey:2012} (at least as long as capacity limits are not taken into account~\cite{Legara:2014}). In addition, the existence of alternative trajectories on different transportation modes enhance the system resilience~\cite{DeDomenico:2014}. 

Inter- and intra-modal connections can be intensively optimized just through modifications and offsetting of the existing timetables, allowing to reduce waiting times at transfer points of a city like Washington D.C. of about $26\%$~\cite{Nair:2012}. A better knowledge of the structure and layout of the Public Transport System would impact a wide range of areas. Indeed, mode choice is one of the fundamental steps in transportation forecasting~\cite{Balmer:2006} and has represented in the past a perfect experimental field for the study of individual choice behavior~\cite{Domencich:1975}. Developments in the availability of urban public transport has the potential to improve significantly the air quality in metropolitan areas~\cite{Meinardi:2008} and directly influences the social geography of a city~\cite{Glaeser:2008}. However, multi-modality not only means the existence of more and better alternative options, but having to deal with all these alternatives at the same time. From the users point of view, the difficulty of dealing with the enormous amount of information needed for describing and taking advantage of the public transport of a city is such that it is no more managed by personal experience and habits, but by services offered by major information technology companies. From the transport agencies point of view, the managing task becomes significantly harder because: i) different modes are run by separate agencies and both data handling and optimization tasks have to cross high organizational barriers; ii) it is not trivial to identify aspects of the system that are relevant for service optimization.

Therefore, in order to help  decision makers, new quantitative approaches are needed to highlight the limits of the system and to assess the impact of new infrastructural development. Our capacity to understand transfer behavior and to evaluate transfer improvements are indeed limited by the lack of proper analytical tools, as we have to take into account many important aspects simultaneously~\cite{Guo:2011}. If we want the public transit system to become a viable alternative to automobile, it is crucial to design cost-competitive and reliable public transportation systems that guarantees both short travel times and a travel experience comparable to those of car trips~\cite{Daganzo:2010}.

Another important difficulty in the study of transportation systems is the data availability. In particular,  it is usually very difficult to obtain traffic related data, and we take advantage here of the availability of another type of data which will enable us to assess the structural efficiency of the system. This open-data information consists in the set of timetables for all transportation modes in the United Kingdom, except for Northern Ireland (see Methods and the Supplementary Information for more details). We will  focus on the urban scale and identify key quantities characterizing the efficiency of the system, providing directions to improve urban Public Transport Systems. More precisely, our goal is to determine:
\begin{itemize}
\item how far is an urban, multimodal public transportation system from optimum,
\item how the temporal aspects impact the structure of quickest paths,
\item how important are the multilayer aspects,
\item the key differences between transportation systems in different cities.
\end{itemize}

Our study is based on the statistical analysis of the quickest paths on the multimodal transportation networks. We assume that origins and destinations are uniformly and independently distributed on the location served by the transport system and we do not take into account access time at the departure location. This uniform demand does not take into account how flows are actually allocated over the network (a piece of information that is usually not directly available) but allows us to focus directly on the structural features of the network, and not on its actual use and on the qualities perceived by the average user. In this sense, the weaknesses and optimization that we discuss here, concern an ideal optimum where all possible routes with all possible origins and destinations would be improved. The methodology developed here can however be very easily adapted to the case where origin-destination matrices are known.

\section{Results}


We first define different types of paths in these systems. In particular, in order to understand how far are urban transportation systems from an ideal optimum, we compare the quickest time-respecting paths with the minimal path. The minimal path is the quickest one, computed by using the largest speed observed on each link and by neglecting waiting times, and represents an unreachable condition, equivalent to having all the existing transportation systems perfectly synchronized for the specific trip under consideration. In contrast, the time-respecting path is the quickest path but where we use the real timetable and where walking and waiting times are taken into account. The time-respecting path is by definition longer than the minimal path and as we can see on an example shown in figure 1, they can be extremely different from each other. In addition, real trips are bound to the transportation system and do not follow a straight line of euclidean distance $d(a,b)$ from the origin $a$ to the destination $b$. The topographical and infrastructural constraints induce differences of the transportation network topology between cities. A consequence of this is that the length $\ell(a,b)$ of the quickest (time-respecting and minimal) trips on the network might be very different from $d(a,b)$, and this difference can be measured by the detour $r(a,b) = \frac{\ell(a,b)}{d(a,b)}-1$ that can be interpreted as a cost-benefit ratio~\cite{Aldous:2010}. In order to compare the availability of routes in different networks, we use the quantity $R = \max_{d>1} r(d)$ for a fixed $d$ subset (see Supplementary Information). The values of $R$ in different cities are strongly anti-correlated (-0.95) with the static network normalized cyclomatic number~\cite{Clark:1991} $M_N = (E-N-1)/N$, reflecting the fact that the more loops are present in the network and the less the detour. In the following, we would like to exclude this topological influence and in order to compare various cities, we will use as a spatial metric the effective length $\ell$ on the network.

Only large cities can afford significant rail-based elements (trains, metro, tram) in their public transport systems and therefore can have a high propensity to interchange~\cite{GUIDE}. Other transportation modes such as ferries and coaches, play a secondary role at an urban level (and air transportation is naturally out of the game). Coaches emerge for minimal paths in certain cities, but their low frequencies are completely excluding them from time-respecting paths. Other forms of road transportation are usually more accessible and, for this reason, bus is the dominant layer for short distances. If cities have enough suitable street space dedicated to Bus and Bus Rapid Transit systems, they are even able to outperform metro and rail systems~\cite{Daganzo:2010}. 

Each transportation mode is characterized by its cruise speed, departure frequencies and accessibility. A consequence of these peculiarities is that, depending on the length of the trip, the public transport system offers different optimal time-respecting solutions. At the national scale (Figure 2, Left), different strategies emerge at different spatial scales. We will not consider very short trips for which the origin and destination are closer than 1km, because distances so short could be easily covered by walking and usually do not rely on transportation systems. Above this scale, the vast majority of short trips are made within the bus layer, and the rail system becomes dominant for inter-urban trips of length larger than approximately $40$ kms. Air transportation emerges naturally for longer distances above $200$kms, and its importance increases significantly for distances of order $400$ kms (e.g. Glasgow-Birmingham) and $d \approx 500$ kms (e.g. Glasgow-London Luton), and becomes finally dominant for trips longer than $700$ kms, connecting for example the southern part of England with the northern part of Scotland.

At the urban level, transportation modes that capture a significant fraction of the time-respecting paths are Bus, Railways and, when available, the Metro and Tramway layer (Figure 3). The bus stops represent the vast majority of the possible origins and destinations and are almost always used in our paths. This bus layer contains in general the largest part of both minimal and time-respecting paths. The use of the fast transportation modes emerges progressively with increasing $\ell$ (Figure 2, right), with a higher rate in larger cities such as London, Manchester and Birmingham, where the Metro-Tramway systems are present. As this transportation mode has a high frequency and a fast speed, and is not affected by congestion, it is naturally used as a quickest alternative to buses across city centres. Nevertheless, due to its limited accessibility, the largest fraction of short trips are done in the bus layer, also in cities where the system has an extended offer of metro lines (Figure 3). The metro layer is in competition with the rail layer, which has higher speed but lower departure frequencies. In cities with high multi-modality (i.e. high average number of modes used per trip), the rail network attracts the largest part of the mobility at distances much lower than at the national level. Indeed, for London (Figure 3, left) and Manchester (Figure 3, right), the length done by train overcomes the one by buses at $\ell \approx 20$ and $\ell \approx 30$ kms.

\subsection{Comparing minimal with Time-Respecting Paths}


We first analyze the multi-modal aspect of trips, quantified by the numbers
$\Lambda_{m(t)}$ which represent the number of different modes for the minimal (m) or time-respecting (t) paths. For some cities, the time-respecting paths display a larger $\Lambda_t$ than for minimal paths, while for others, it is the opposite (Fig.4a-b). The relative loss in multi-modality due to synchronization can be measured by
\begin{equation}
\Delta = \frac{\langle \Lambda_m-\Lambda_t\rangle}{\langle \Lambda_m -1\rangle}
\end{equation}
The larger $\Delta$ and the larger the difference between minimal and time-respecting paths. We see in Fig. 4c that $\Delta$ is positive when the average speed of the alternative (ie. non-bus) layers $V_{nb}$ is sensibly larger ($>2.5$ times) than the average speed $V_b$ of the bus layer (see Supplementary Information for more details). The quicker the rail and metro layers are, the more multimodal the minimal path would tend to be. Indeed, for minimal path, the use of fast non-bus layers is only limited by their accessibility, i.e. by the extra-time needed for reaching the inter-layer connection point. For time-respecting paths, multi-modality also implies the importance of synchronization, and it appears that in cities where metro or rail are sensibly faster, their frequency is also lower (Fig. 4d). In other words, in cities where the fast layers are extremely advantageous in term of speed, the system suffers from synchronization problems.  This empirical finding suggests the existence of a structural limit to transportation systems' possibilities that policy makers should take into account in the search of a efficient optimization strategy.

As a consequence, if the rail and metro layers are relatively quick, they are used for minimal paths while additional waiting times due to mode change can be too costly for the time-respecting paths. On the other hand, in cities where the bus layer is fast but with a frequency as low as for faster layers (eg. London, Liverpool, Cardiff), minimal paths tend to use buses only, while the time-respecting paths face the synchronization limits of the bus layer itself (see for example figure 1). More generally, the factors responsible for the time difference between time-respecting and minimal paths are: (i) waiting times (both intra- and inter-layer);  (ii) the fact that the optimal riding times used to compute minimal paths may differ from the riding times at a particular hour; (iii) a long walking time for connecting different modes in a wide stop area. In order to quantify the differences between minimal and time-respecting paths, we introduce the synchronization inefficiency $\delta$, computed as the ratio of time-respecting travel time $\tau_t$ and minimal travel time $\tau_m $
\begin{equation}
\delta =\frac{\tau_t}{\tau_m} - 1
\end{equation}
For all cities, $\delta$ reaches its maximum $\delta_{max}$ for short trips, where waiting times are long compared to the travel time, and then decreases with the distance $\ell$ according to the following function, valid for all cities (see Fig. 5, left)
\begin{equation}
\delta \approx \delta_{min} + \frac{\delta_{max}-\delta_{min}}{\ell^\nu}
\label{deltafun}
\end{equation}
where $\nu\approx 0.5$. The collapse observed for $\delta$ for all cities suggests that there is an underlying process describing the accumulation of waiting and walking times along time-respecting paths. The specifics of the different cities appear in the system efficiency in both the worst $\delta_{max}$ and best $\delta_{min}$ limits. We note here that this Eq.~\ref{deltafun} is consistent with a simple argument based on the central limit theorem leading to $\nu=1/2$.

Time-respecting paths are however not completely different from the minimal ones and we can measure the similarity of two paths by using their spatial overlap $q$, defined as the fraction of length of edges they have in common. The overlap is $q=1$ for extremely short trips (if a single edge is used, waiting times are not playing any role), and then decreases with $\ell$ for all cities as (Fig 5, right)
\begin{equation}
q \approx q_{min} + (1-q_{min})e^{-\ell/\ell_q}
\label{qfun}
\end{equation}
where  $\ell_q$ is the scale parameter for each city. The function $q(\ell)$ converges to a limiting value $q_{min}$ in the range $[0.15,0.33]$. This minimal overlap is due to the limited number of good options available, especially close to the origin and the destination. The constraints due to the local connectivity and optimal cruise speeds make the minimal path the best option also when  time causality starts playing a role. 
The exponential decay of the overlap with $\ell$ suggests that there is a typical `branching' length $\ell_q$ for each city, 
which sets the probability of having alternate routes. In other words, the probability for the time-respecting path to deviate at each $d\ell$  from the minimal path is proportional to $\ell_q^{-1}$.

\subsection{Anatomy of a trip}

We have seen so far that the length $\ell$ governs the behavior of most quantities characterizing a trip. In order to identify the role of the temporal and multilayer aspects of the network in the structure of the time-respecting paths, we detail how the total travel time can be decomposed into different components: the riding time (with any mode), and waiting and walking times at interchanges. In addition, in order to take into account the multi-modal aspect of the network, we discriminate riding times per layer and we separate intra-layer from inter-layer waiting time. This wide spectrum of temporal quantities forms what we call the `anatomy' of a trip, and is represented in Figure 6 a-c for different cities. This figure allows for a quick understanding of how the temporal structure of trips varies with distance. 

We first note that the travel speed grows with $\ell$ (see Supplementary Information) which implies that travel time for time-respecting paths grows sub-linearly with the distance covered. Another important contribution in trips is due to walking between modes in the multilayer network, which represents a fixed cost of multi-modality (in addition to inter-layer waiting times). We naturally expect walking times to grow with $\ell$ as the number $\Lambda_t$ of layers used in time-respecting paths (see Supplementary Information for more details). Finally, waiting times are one of the two main contributions to the synchronization inefficiency $\delta$, the other being the difference between optimal and actual edges' riding times. These times will play a relatively minor role for long distances, as their relative importance compared to riding times decreases.

The analysis of these anatomy plots shows the following. At short distance, in all cities but London, most of the travel time is spent in intra-layer waiting time. Most trips start at the bus layer, and the first connections within this layer are those that make the system extremely inefficient. We therefore define for each city a distance $\ell_w$ such that for trips shorter than $\ell_{w}$ the waiting time represents more than $50\%$ of the travel time. For $0<\ell<\ell_w$, the lack of synchronization is dominant and the temporal network is far from being optimal. This distance interval corresponds to values of $\ell$ where the overlap $q$ is larger than $50\%$ (Fig. 6d). For short trips, we have  few alternative paths and cannot avoid waiting times due to synchronization problems. Time-respecting paths are thus very similar to minimal paths, and (large) waiting times are directly added to optimal riding times.

For long distances, we already saw that the multi-modal nature of the systems becomes important. The use of fast transportation modes becomes advantageous only when the difference in speed compensates for the time necessary to reach the rail or metro network. In order to measure this effect, we define the distance $\ell_{nb}$ such that for $\ell>\ell_{nb}$ the largest part of the trip is done with a transportation mode different from the bus. Trips with $\ell_{w}<\ell<\ell_{nb}$ are then essentially made within the bus layer, and most of the travel time is due to actual transfer (walking or riding). We observe for large cities like London, Manchester, or Birmingham, a finite value of $\ell_{nb}$ indicating that at a certain point the bus layer loses its dominant role. This is in contrast with smaller cities where $\ell_{nb}$ is larger than the city radius, implying that fast layers always play a marginal role in these cases.

If we take into account time respecting paths with at least one inter-layer connections only, we find that on average for all cities considered in this work, the time spent in connections (walking and inter-layer waiting times) represents a significant fraction ($23\pm6 \%$, see Supplementary Information) of the total travel time. The different regimes identified in figure 6d suggest that different strategies might have the better impact for each city for optimizing the transport time of trips of specific distances. Short trips are indeed dominated by intra-layer waiting times, while long trips by riding times. In the cases where the multimodality becomes dominant, inter-layer waiting and walking times, together with the fast layers' cruise speed, become instead the most relevant quantities for the optimization task.

\subsubsection{The role of the total number of stop events}

An interesting question concerns the characterization of a multilayer, temporal network such as the transportation systems
that we consider here. Obviously, the number of modes and their frequency play important roles in their efficiency. A simple, natural quantity is then given by the average number of stop events per unit time 
\begin{equation}
\Omega = \frac{\sum_\alpha C_\alpha}{\Delta t}
\end{equation}
where $C_\alpha$ is the number of stop events in the layer $\alpha$ and $\Delta t$ the duration of the time interval considered in the analysis of the temporal network (see Methods). In this study, we considered a starting time of 8:00 am (monday) and a duration $\Delta t=16h$ which covers a whole day of mobility. The quantity $\Omega$ represents a global measure of the transportation service offered in a city, of the infrastructural cost of the transport network, and is indeed proportional to the cities population (Fig 7a). In order to improve the transportation system and to serve more people, one may add new lines, new connections, increase the frequency of a line, or even introduce a new transportation mode in the network, and the quantity $\Omega$ integrates all these modifications. 

As we will see, it is actually remarkable and unexpected that a single network indicator such as $\Omega$  is enough to explain the behavior of many key quantities characterizing the public transport network of different cities. For example, the interplay between temporal and multilayer aspect of the public transport network is highlighted in Fig 7b, showing that the fraction $\lambda_t$ of time-respecting paths  using more than one mode~\cite{Morris:2012} is larger for cities with a larger number of stop events. If we assume that the average number of possible alternative to bus layer path (which is always an available option) is $a_{\bar\lambda} \Omega$, the expected fraction of unimodal trips is $\omega(\Omega) = (1+a_{\bar\lambda}\Omega)^{-1}$. The average interdependency of the time-respecting paths is then $1-\omega$
\begin{equation}
\bar\lambda_t \approx  \frac{a_{\bar\lambda} \Omega}{1+a_{\bar\lambda} \Omega}
\label{eq:lamb}
\end{equation}
Using this form to fit the data shown in Fig. 7b, we obtain $a_{\bar\lambda}=1.65\ 10^{-5}$hour/stops

Similarly, $\Omega$ is related to the cruise speed $V_{cruise} = \ell/\tau_{cruise}$ (where $\tau_{cruise}$ is the time spent in a moving vehicle) and therefore also to the time respecting paths travel speed $V_{travel} = \ell/\tau_t$ (Fig.7c). Indeed, we can assume that the fraction $\omega$ of unimodal paths is traveled at the average bus layer speed $\bar V_b$, while the fraction $1-\omega$ of multi-modal paths is traveled at a speed $\bar V_{multi}=\frac{\bar V_b + k\bar V_{nb}}{1+k}$,  where $k=\ell_{nb}/\ell_b$ represents the ratio of lengths on the non-bus and bus layers ($\bar V_{nb}$ is the average speed for the fast layers --- see Supplementary Information). The average travel speed $\bar V_{travel}$ grows then with $\Omega$ as
\begin{equation}
\bar V_{travel,th} = (\omega(\Omega)\bar V_b+(1-\omega(\Omega)) \bar V_{multi})\frac{\tau_{cruise}}{\tau_{t}}
\label{vtravel}
\end{equation}
Using the value for $a_{\bar\lambda}$ obtained above, we minimize the variance between the estimated and empirical values of $\bar V_{travel}$ and we find the optimal value: $k\approx 0.8$ (see Fig. 7c). This result shows that for all cities considered here (and under uniform demand), approximately $45\%$ of time-respecting paths are on non-bus transportation modes.

In addition, the quantity $\Omega$ also characterizes the efficiency of a public transportation network in terms of synchronization. Indeed, we observe that the average synchronization inefficiency measure $\bar\delta$ decreases with $\Omega$ as a power law (see Fig. 7d)
\begin{equation}
\bar\delta \approx  \Omega^{-\mu}
\label{eq:deltaomega}
\end{equation}
where $\mu\approx 0.3\pm 0.1$.  The expected decrease is naturally due to the fact that larger values of $\Omega$ implies larger frequency and thus a better synchronization between modes. The small value of $\mu$ is however bad news in terms of efficiency: in order to divide $\overline{\delta}$ by a factor $2$ we need to multiply $\Omega$ by a factor of almost $10$. We can however hope that when exact origin-destination matrices are known, a better optimization of the system can be obtained through targeted improvements. It is not unusual to observe power law behavior in urban systems~\cite{Bettencourt:2007}\cite{Louf:2014}, and although the fits are not perfect (essentially due to the small number of available decades), this result Eq.~\ref{eq:deltaomega} could be useful for constructing coarse-grained models of transportation in cities.  Besides this, we note here that the city of Edinburgh is an outlier in all figures 7(a-d) and, for this reason, has been excluded from the best-fit of figures 7b and 7c. Indeed, even if $\Omega$ is relatively high for this city, Edinburgh's public transport system seems to use a significantly different strategy in managing the mobility demand, characterized by an extremely high bus-frequency. The network is therefore extremely efficient in terms of synchronization but not performant in terms of cruise speed (see fig 4d), as can be seen with time-respecting paths that are mostly composed of slow unimodal bus trajectories (see figures 7b and 7c).

\section{Discussion}

We identified the total number of stop events per hour $\Omega$ as the key quantity which characterizes the efficiency of a transportation system and its efficiency in terms of speed, multimodality and synchronization. Naturally, $\Omega$ is not the only parameter at play: multi-modality depends also on the different cruise speed and departure frequencies in the different layer. In the UK transportation system, these two quantities are anti-correlated, as if a city system might try to optimize rail and metro systems, with respect of the bus system, either making them faster or more frequent. This relationship has important practical applications, as it constitute a limit that policy makers need to take into account in their system optimization, and can serve as a support for evaluating alternative Public Transport Systems' designs.

The temporal aspect of the Public Transport Networks appears to be influential for trips covering all distances. Short time-respecting paths tend to be mostly similar to the minimal ones, and waiting times are directly added to the riding times of the associated minimal paths. Waiting times then represent the largest fraction of the total travel times, and at this scale an increase of bus departure frequency, or methods like timetables offsetting~\cite{Coffey:2012, Nair:2012} of the bus service may represent a good optimization strategy. Longer time-respecting paths tend instead to diverge from minimal ones, and very large waiting times can be avoided thanks to the availability of alternative routes, and when it is possible, longer trips are progressively taking advantage of the multi-modality of the system. For cities with a large level of multi-modality, as it is the case for London, Birmingham Manchester, it becomes hard to disentangle the temporal and multilayer aspects of the system. Waiting time (together with walking time) does not represent a simple cost to minimize, but a price to pay to access to fast transportation. 

The value of waiting and walking times are perceived as higher than the time spent travelling~\cite{Wardman:2004}, in particular because walking demands a greater physical effort~\cite{Kolbl:2003}. Waiting time has an higher perceived cost because of the frustration due to the sheer inconvenience of waiting~\cite{Wardman:2004}. All these costs have to be integrated with those related to the time needed for accessing the network~\cite{Daganzo:2010}, the stress of the transfer experience~\cite{Guo:2011},  breaking personal habits~\cite{Garling:2003}, scheduling costs and those caused by the unreliability of arrival times~\cite{Wardman:2004}. In order to optimize the travel experience and to minimize the perceived mobility cost, it is then necessary to consider the full anatomy of trips and to distinguish between transportation modes and between the nature of time spent (riding, waiting, walking). In this respect, we believe that the tools and the methodology developed here will allow for an integrated view of these systems and will be helpful for testing and finding specific optimization strategies.

\section{Methods}

\subsection{Data} 
\label{sectionData}
The land transport timetables used in these papers are provided by the National Public Transport Data Repository~\cite{NPTDR} under Open Government licence. A snapshot of every public transport journey is recorded for all services running in Great Britain (England, Scotland, Wales) during a full week in October 2010.  The raw files contain the information available in the travel-lines web sites and call-centres during the selected week. For road transport, transportation agencies take into account the average traffic conditions at different hours and days for the design of timetables, so that they implicitly contain congestion effects. 

The modes covered and identified are bus, coach, train (national rail), ferry and metro (including Underground, tram, light rail and non-national rail trains). All routes are referenced to stops coded using the NaPTAN scheme (National Public Transport Access Nodes) data~\cite{NaPTAN}. In the NaPTAN scheme, every UK rail or metro station, coach terminus, airport, ferry terminal, bus stop or taxi rank is associated to at least one Stop Point. Not all Stop Points are actually used, so only those that were present in the timetables are considered active and have been taken into account. Stop point are then organized in Stop Areas representing facilities (Airports, Bus/Metro/Coach/Railway Stations) or possible interchange points. The definition of these Stop areas has been taken as a basis for defining a multilayer network from the timetable data. A further process of data cleaning and aggregation has been performed to have a consistent definition of inter-modal exchange points (see Supplementary Information).
To complete the spectrum of transportation modes, we use detailed schedules of all non-stop UK domestic flights, provided by Innovata LLC~\cite{innovata} for the week of 18-24 October 2010. Each of these flights has been associated to the Stop Points of the arrival and departure airport (and eventually to a specific terminal). The multilayer temporal network dataset derived from these data is publicly available at {\bf http://www.quanturb.com/data.html }

\subsection{Multilayer temporal network}
The inter-modal exchange points are identified by (i) original NaPTAN Stop Areas, (ii) new Stop Areas obtained by a spatial aggregation of Stop Points (see Supplementary Information). To be an exchange point, journeys of different transportation modes should stop in that Area and to correctly define a Multilayer network~\cite{Kivela:2014}, we associate all Stop Points to a layer $\alpha$, representing a specific transportation mode. If a Stop Point belongs to a Stop Area, the point is not represented in the network and all vehicle stops in that point are associated to the area. Both Areas and Points are identified by an id $i$. As buses and coaches may stop in the same location, a copy of the same Stop Point can be defined in two different layers, and thus associated to two different vertices $v_{i\alpha}$ and $v_{i\beta}$ in the multilayer network. Similarly, if an Area $i$ has associated points belonging to a set of layers ${\alpha,\beta,\gamma,\dots}$, a vertex representing that Area is defined in each of those layers ($v_{i\alpha}$, $v_{i\beta}$, $v_{i\gamma}$, \dots). Inter-layer edges connects all couples of vertices associated to the same Point or Area in different layers in both directions $(i_\alpha,i_\beta)$ and  $(i_\beta,i_\alpha)$. If the connection from a layer $\alpha$ to a layer $\beta$ is performed by walking, a walking distance is assigned to each of these edges $(i_\alpha,i_\beta)$ which is calculated as the average distance between all couples of active Stops Points in $i$ belonging to the two different layers $\alpha$ and $\beta$. The travel time has been then computed using a standard walking speed of 5 km/h~\cite{Daganzo:2010}. In addition to the walking times, additional 30 minutes are added to the inter-links from the air-flights layer to all the others, in order to take into account the characteristic waiting times in airports. Similarly, two hours of check-in and security control times are added to the inter-links towards the airline layer (which corresponds to the time suggested by airlines to be at the airport before departure time).

In each layer $\alpha$, we thus have a set of $N_\alpha$ vertices, representing stops locations. The timetables define a set of events occurring in these vertices. Each vehicle departure can be associated to a directed connection between two vertices  $v_{i\alpha}$ and $v_{j\alpha}$ that occurs at a certain time. These events can be represented as $C_\alpha$ quadruplets $(i,j,t,\delta t)$, where $i,j\in V_\alpha$, $t$ denotes the departure time and $\delta t$ the riding time for that specific trip~\cite{Pan:2011}. Besides the temporal network, we can also study the static topology of the public transport network by defining a set of $E$ edges, where the edge $(i_\alpha,j_\alpha)$ exists if there is at any time at least a connection between $v_{i\alpha}$ and $v_{j\alpha}$. For each of these edges, we compute the minimal riding time observed at any time $\delta t_{min}$. We define the minimal path as the shortest path on this static network, where the cost associated to each link is the minimal riding times. We use these minimal paths as a benchmark which represents the optimal mobility through the multi layer network. All the measures performed in this paper are limited to the largest strongly connected component~\cite{Clark:1991} of the static network associated to the corresponding area.

\subsection{Time-Respecting Paths} 
Paths performed through the network must respect the time-ordered sequences of contacts. For this reason, a journey has to follow causal temporal paths defined as a sequence of connections with non-decreasing times \cite{Holme:2012}. We define the travel duration $\tau_{ab}(t)$ as the shortest time needed to reach $b$ starting from a connection from $a$ departing at a time $t' \geq t$. The duration is not static but depends upon $t$. In this paper, we  focus on the morning rush hour, and thus we chose $t_0 =$ Monday, 8:00 am. The temporal distance is measured starting from the actual beginning of the trip, without taking into account the first waiting time $t'-t$. Furthermore, to limit the contribution of a small number of location from where connections are extremely rare, we introduce a waiting time cutoff $\Delta_c=2$h limiting the maximum delay allowed for a single connection~\cite{Pan:2011}. Even while working on a static connected component, this cutoff limits the number of allowed paths. At a national scale, approximately $16\%$ of the trips in the largest strongly connected component of the static network have been excluded because unreachable with this choice of $t_0$ and $\Delta_c$.

\section{Acknowledgments} 
The authors are supported by the European Commission FET-Proactive project PLEXMATH (Grant No. 317614)

\section{Author Contributions Statement}
RG, MB designed, performed research and wrote the paper.

\bibliographystyle{prsty}



\begin{figure*}
\centerline{
\includegraphics[angle=0, width=.7\textwidth]{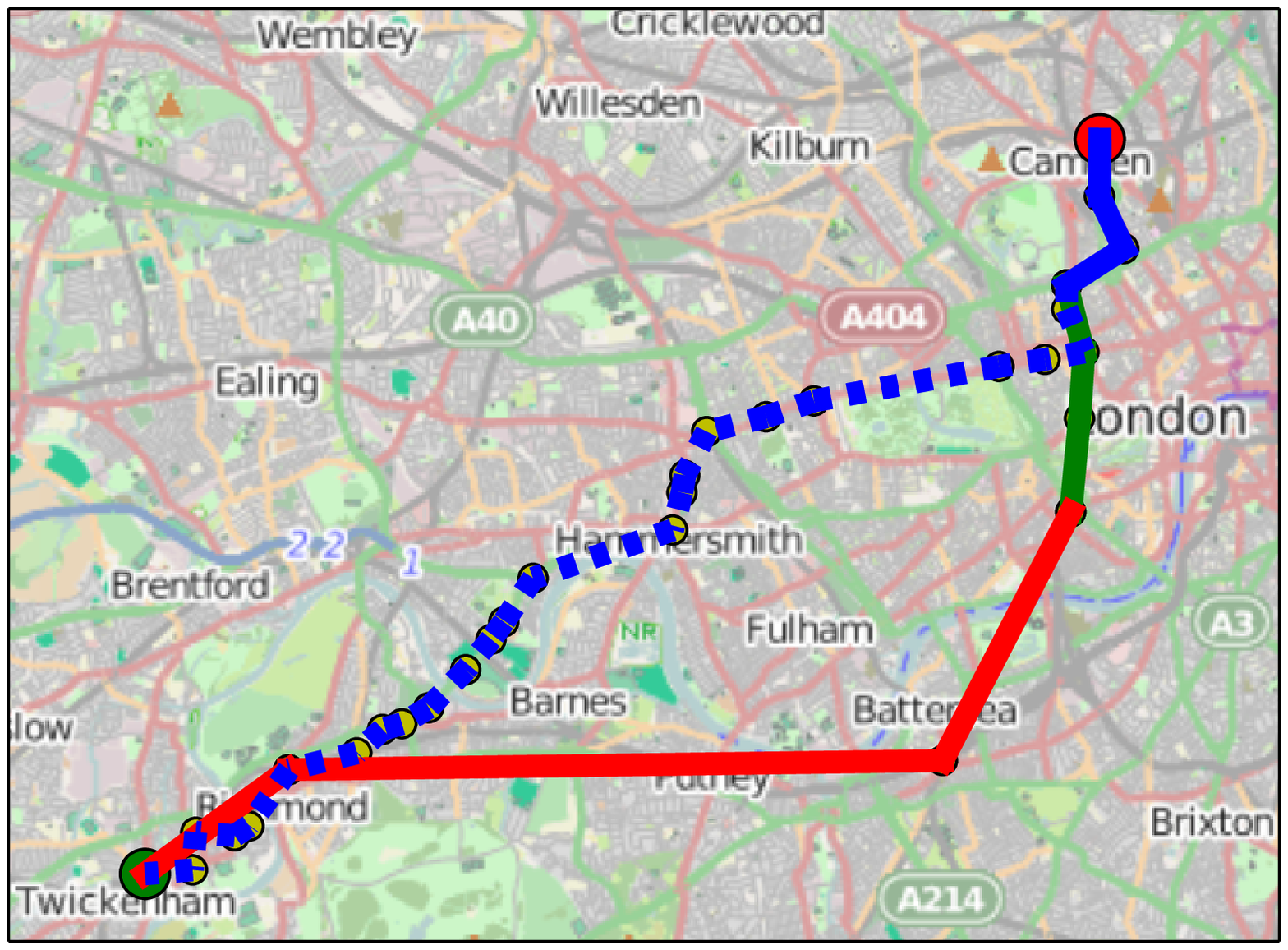} 
}
\caption{Example of the difference between a time-respecting path (solid line) and a minimal path (dashed line). Here, we show a trip from Twickenham to Camden Town in London, and the minimal path would use only buses (marked in Blue). The optimal time-respecting path in this case is remarkably multi-modal: the bus layer is still used for the final segment, which is the same as for the minimal path, while the Rail layer (Red) and then the Metro layer (Green) are used for approaching the city center. [Figure created with Basemap Matplotlib Toolkit for Python using map tiles from openstreetmap.org (\textcopyright OpenStreetMap contributors\cite{openstreetmap}, licensed as CC BY-SA).]}
\end{figure*}


\begin{figure*}
\centerline{
\includegraphics[width=0.45\linewidth]{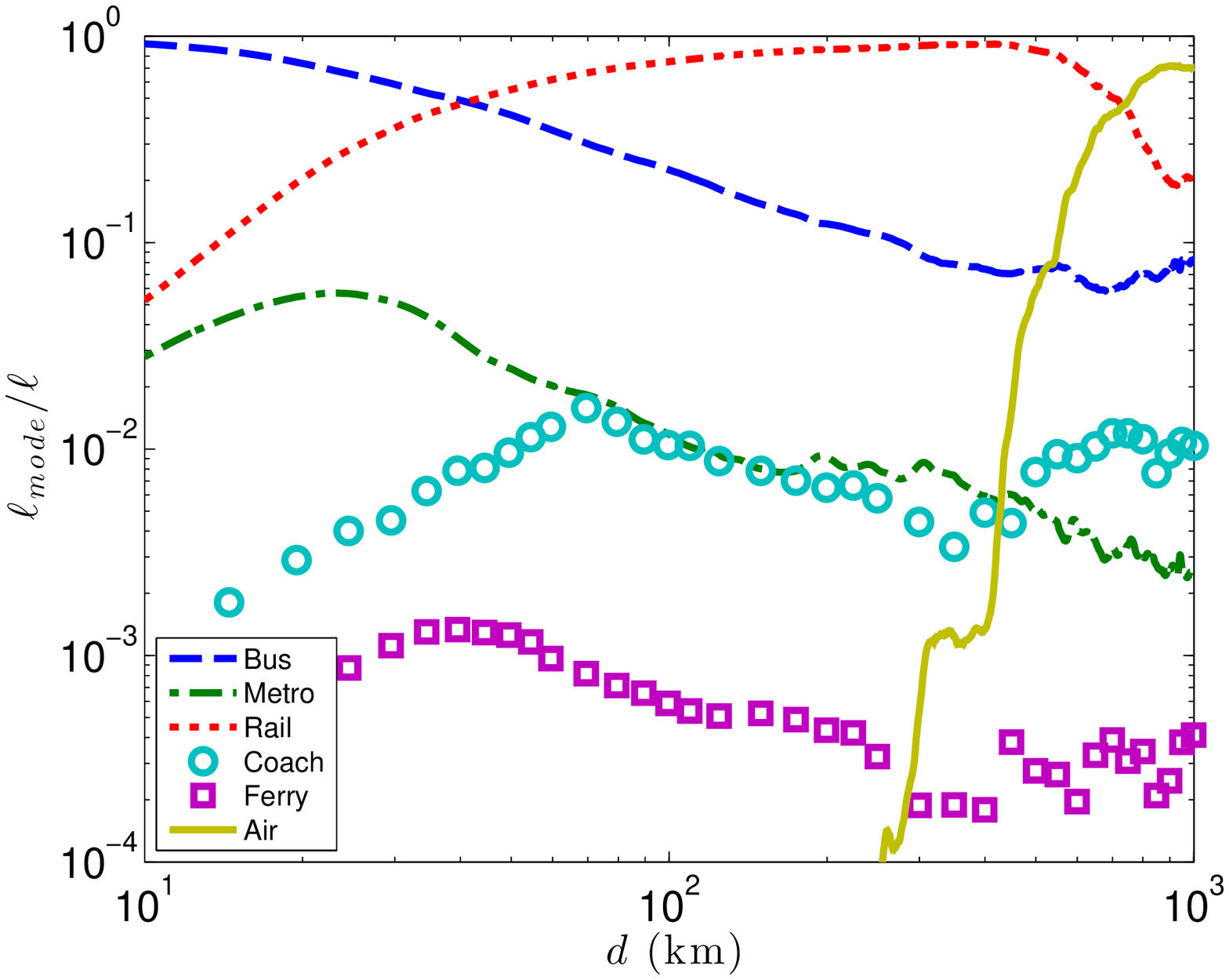}
\quad
\includegraphics[width=0.45\linewidth]{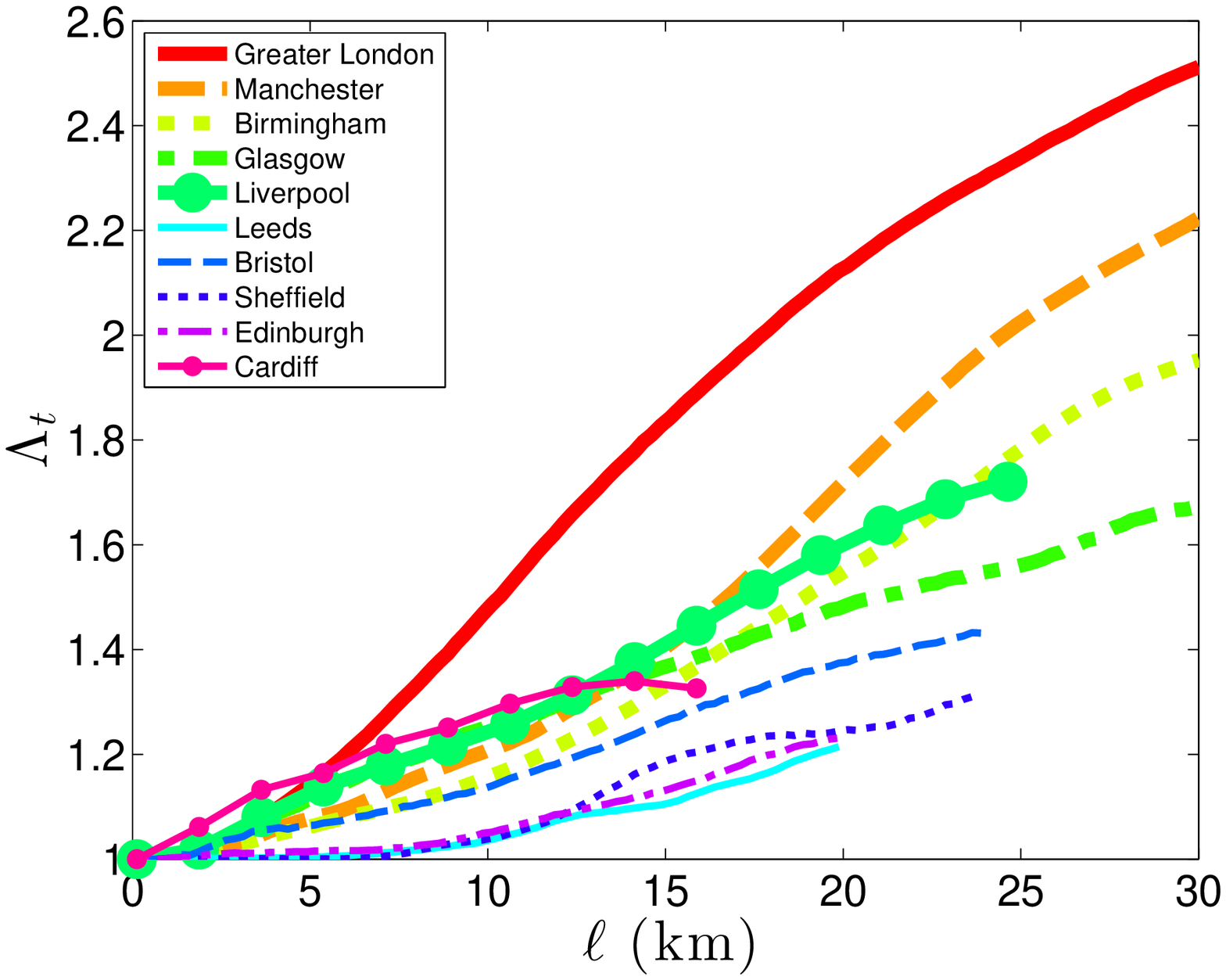}
}
\caption{(Left) Fraction of distance covered by the different modes for time-respecting paths through the whole Great Britain. Short trips are mostly done by bus. Rail becomes then dominant at $40$kms and air travel is dominant for trips of distance of order $700$kms. Other transportation modes play a secondary role, with peaks at  $22$kms for the Metro, $40$kms for Ferries and $70$kms for Coaches (which are increasingly used for long distances as a mean to connect to airports). (Right) The number of modes $\Lambda_t$ used in time-respecting paths through urban areas versus the trip length $\ell$. Larger cities (London, Manchester, Birmingham) show a particularly marked trend towards multi-modality with an average of more than 2 modes for trips longer than $20$kms.}
\end{figure*}


\begin{figure*}
\begin{center}
\includegraphics[angle=0, width=0.45\textwidth]{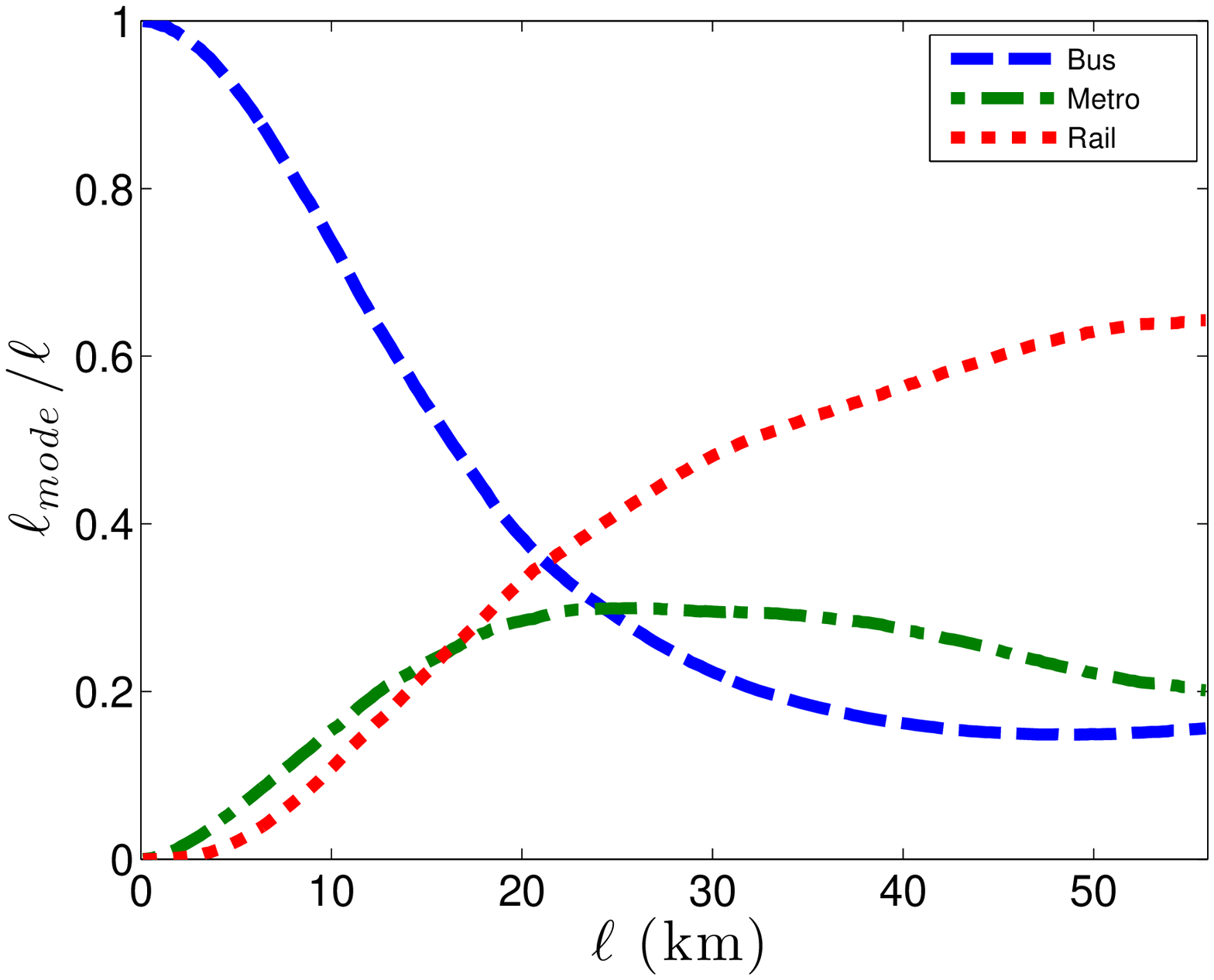}
\quad
\includegraphics[angle=0, width=0.45\textwidth]{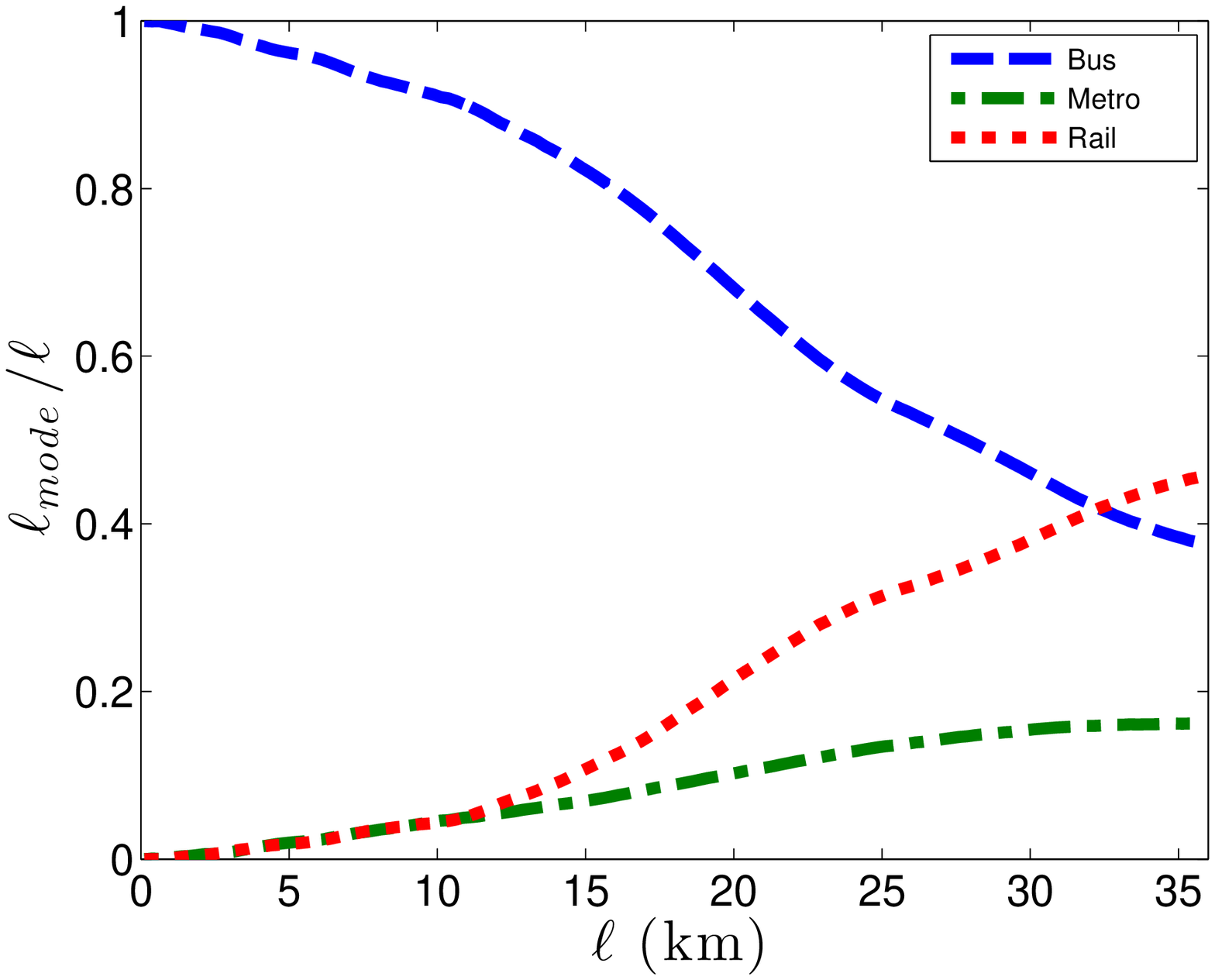}
\end{center}
\caption{Fraction of distance covered with different modes for time-respecting paths through London (Left) and Manchester (Right). The bus system is covering most of the short trips, whereas the advantage of using the Metro and Rail systems emerges progressively for longer distances. Metro networks were naturally developed for answering urban transportation demand and we see that its use competes with the rail system for distances shorter than $15$kms (for larger distances, rail prevails).}
\end{figure*}


\begin{figure*}
\begin{center}
\begin{tabular}{cc}
\raisebox{2.3cm}{(a)} \includegraphics[angle=0, width=0.45\textwidth]{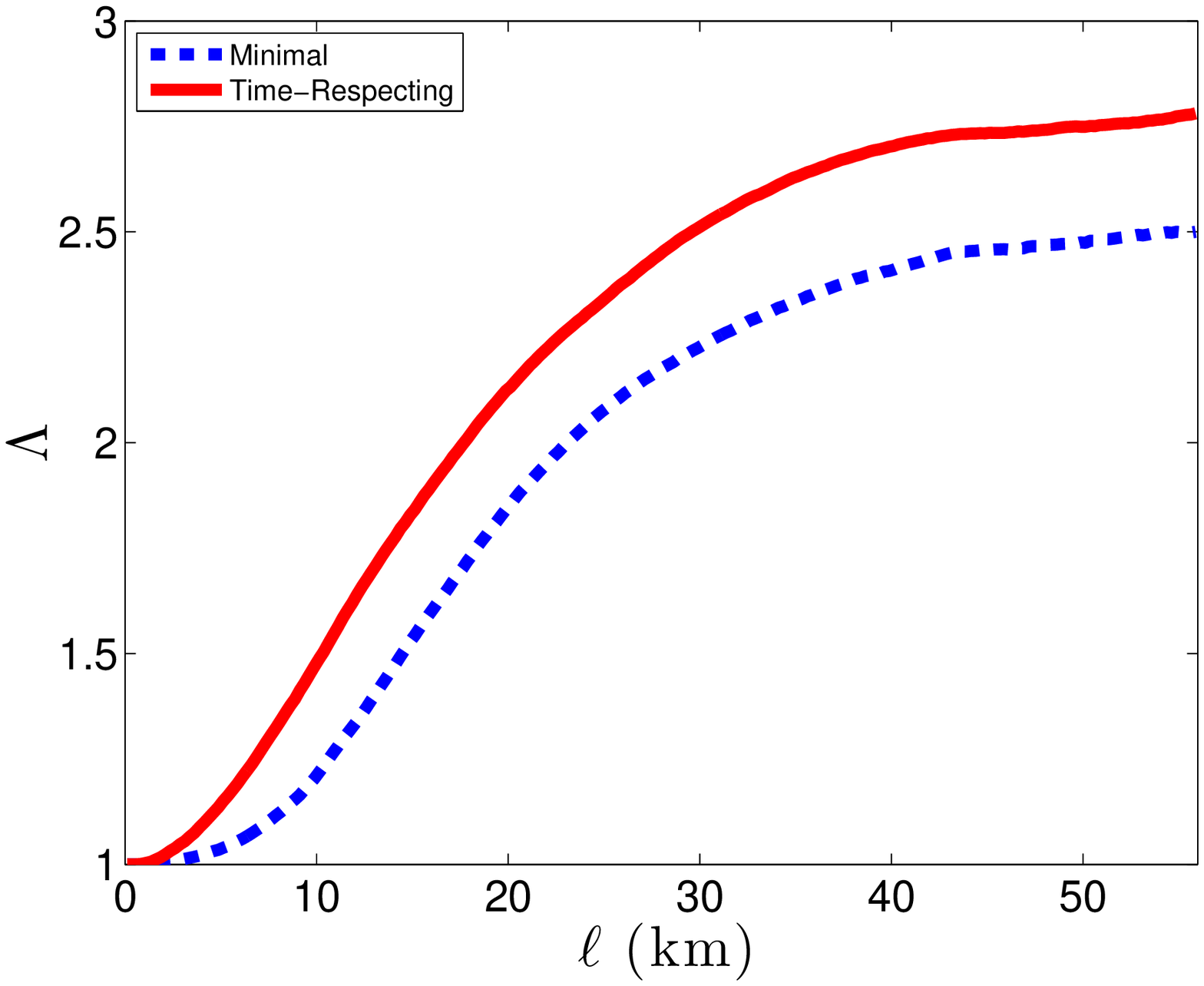}&
\raisebox{2.3cm}{(b)} \includegraphics[angle=0, width=0.45\textwidth]{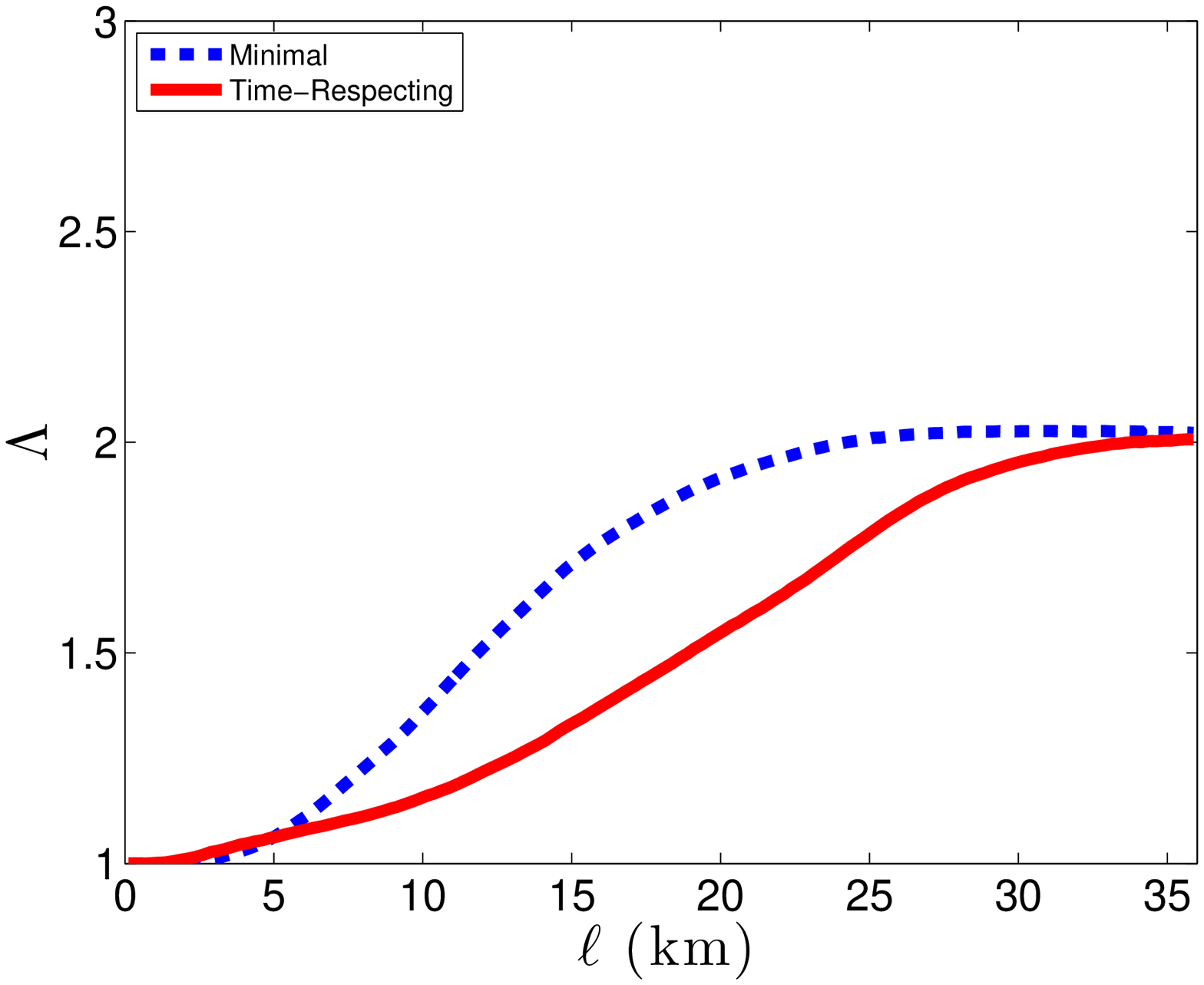} \\
\raisebox{2.3cm}{(c)} \includegraphics[angle=0, width=0.45\textwidth]{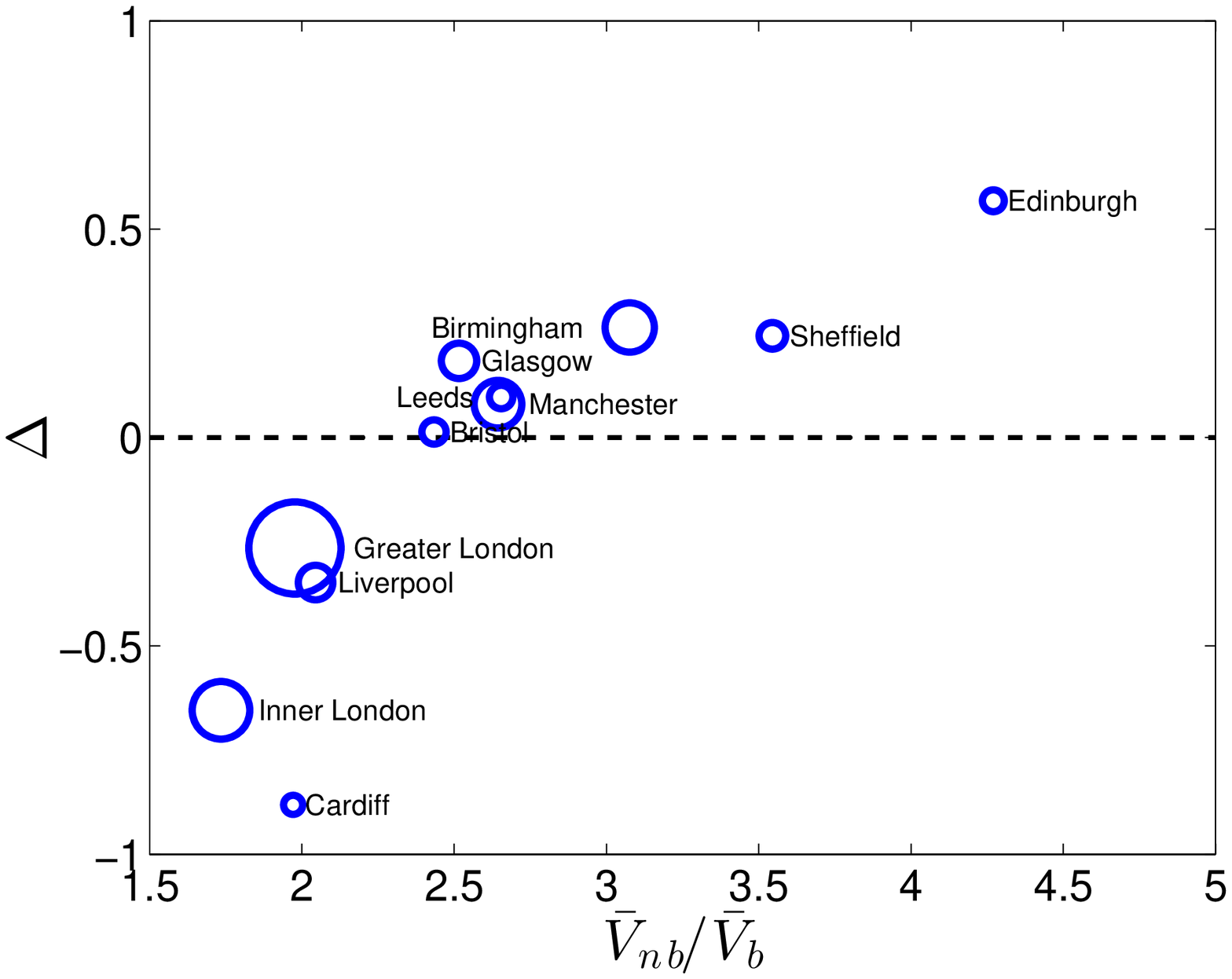} &
\raisebox{2.3cm}{(d)} \includegraphics[angle=0, width=0.45\textwidth]{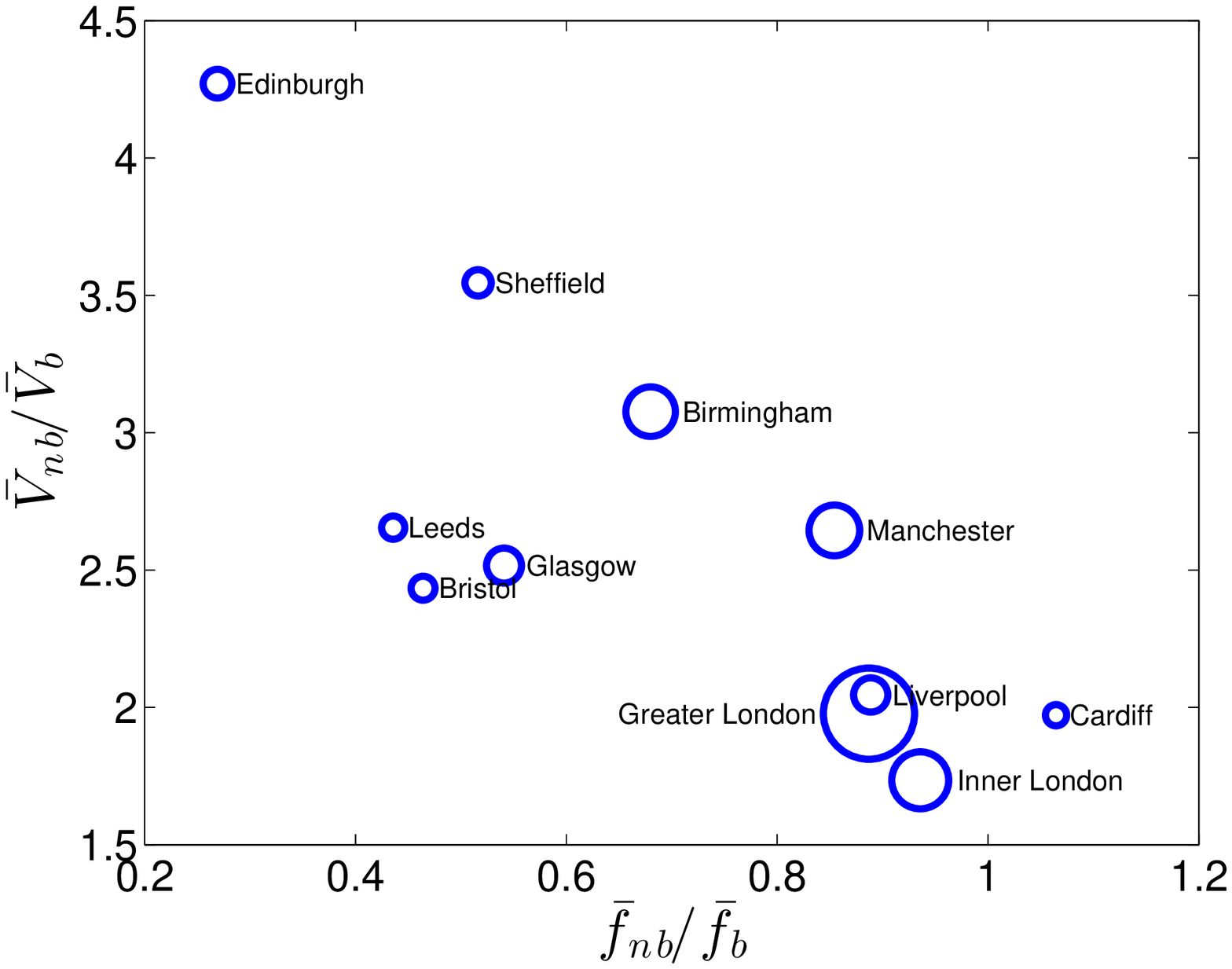} \\
\end{tabular}
\end{center}
\caption{Time-respecting paths may be more multi-modal than minimal paths. As shown with the average number of modes $\Lambda$, it indeed is the case in London (a), while in Birmingham (b) this happens only for extremely short trips. (c) The loss in multi-modality, due to synchronization in time-respecting path, is related to this speed advantage $V_{nb}/V_{b}$ from using non-bus layers instead of the bus layer: when the rail or metro layer is fast, it also usually suffers from synchronization problems. (d) The speed advantage $V_{nb}/V_{b}$ is limited by the frequency advantage $f_{nb}/f_{b}$: even if the Rail layer has faster cruise speeds, its lower frequency implies larger waiting times. We find that for average values computed for different cites, these two quantities are anti-correlated. This relationship suggests that the competition between speed and frequencies is possibly structural and intrinsic to transportation networks. In (c) and (d) the circle's areas are proportional to the city's population. 
}
\end{figure*}


\begin{figure*}
\centerline{
\includegraphics[angle=0, width=0.45\textwidth]{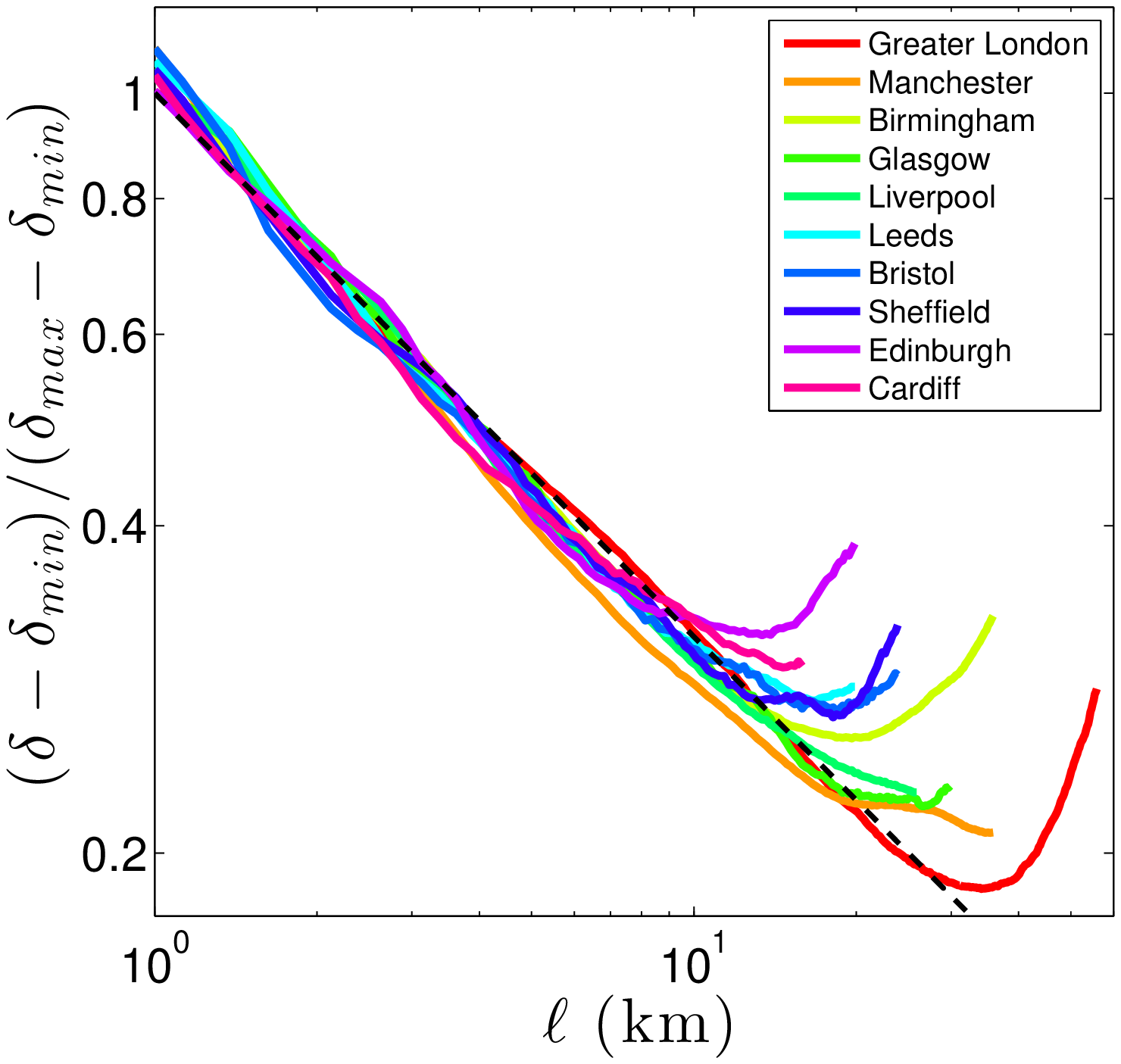}
\quad
\includegraphics[angle=0, width=0.45\textwidth]{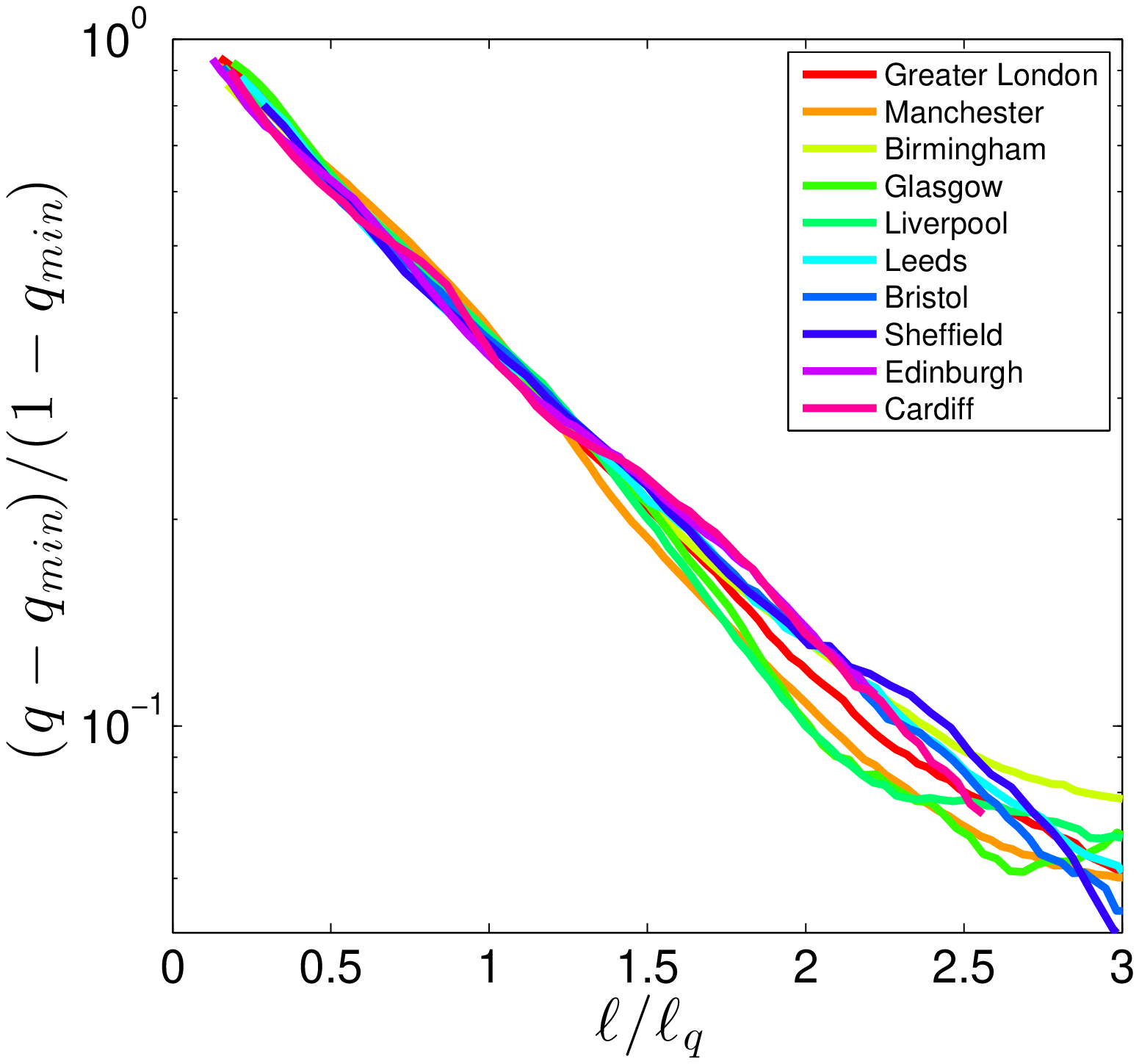}
} 
\caption{(Left) Dependance of the synchronization inefficiency  $\delta$  on the path length $\ell$. For all cities, we observe the same function Eq. (\ref{deltafun}), where the specificity of each city is seen in the values of the peak $\delta_{max}$ and minimal $\delta_{min}$. (Right) Overlap $q$ between minimal and time-respecting paths versus path length $\ell$. The overlap decreases exponentially with $\ell$ to an asymptotic non-zero value $q_{min}$ (eq. (\ref{qfun})). The values of $q_{min}$ and the characteristic length $\ell_q$ encode the differences between the cities transportation networks.
}
\end{figure*}


\begin{figure*}
\begin{center}
\begin{tabular}{cc}
\raisebox{2.3cm}{(a)} \includegraphics[angle=0, width=0.45\textwidth]{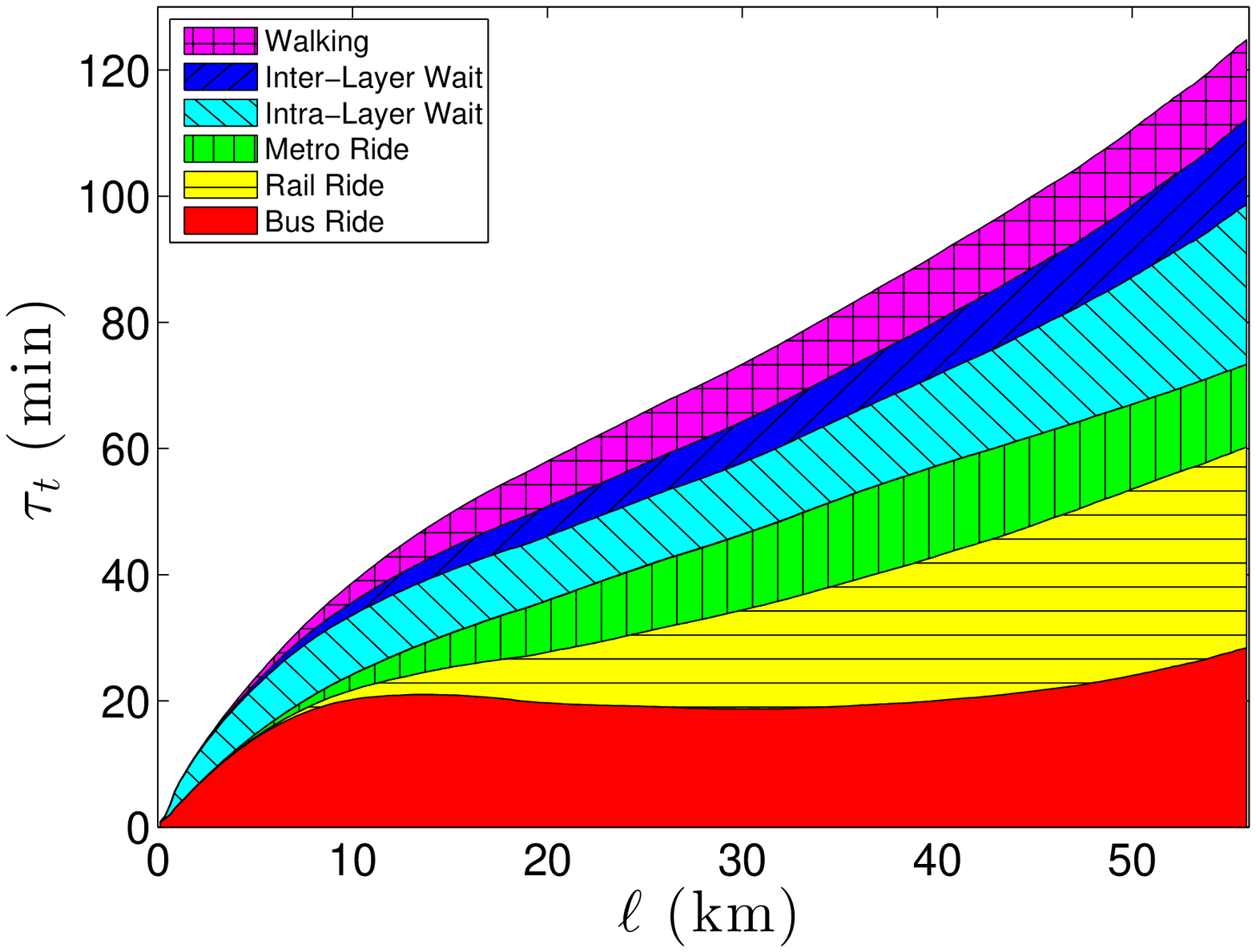}&
\raisebox{2.3cm}{(b)} \includegraphics[angle=0, width=0.45\textwidth]{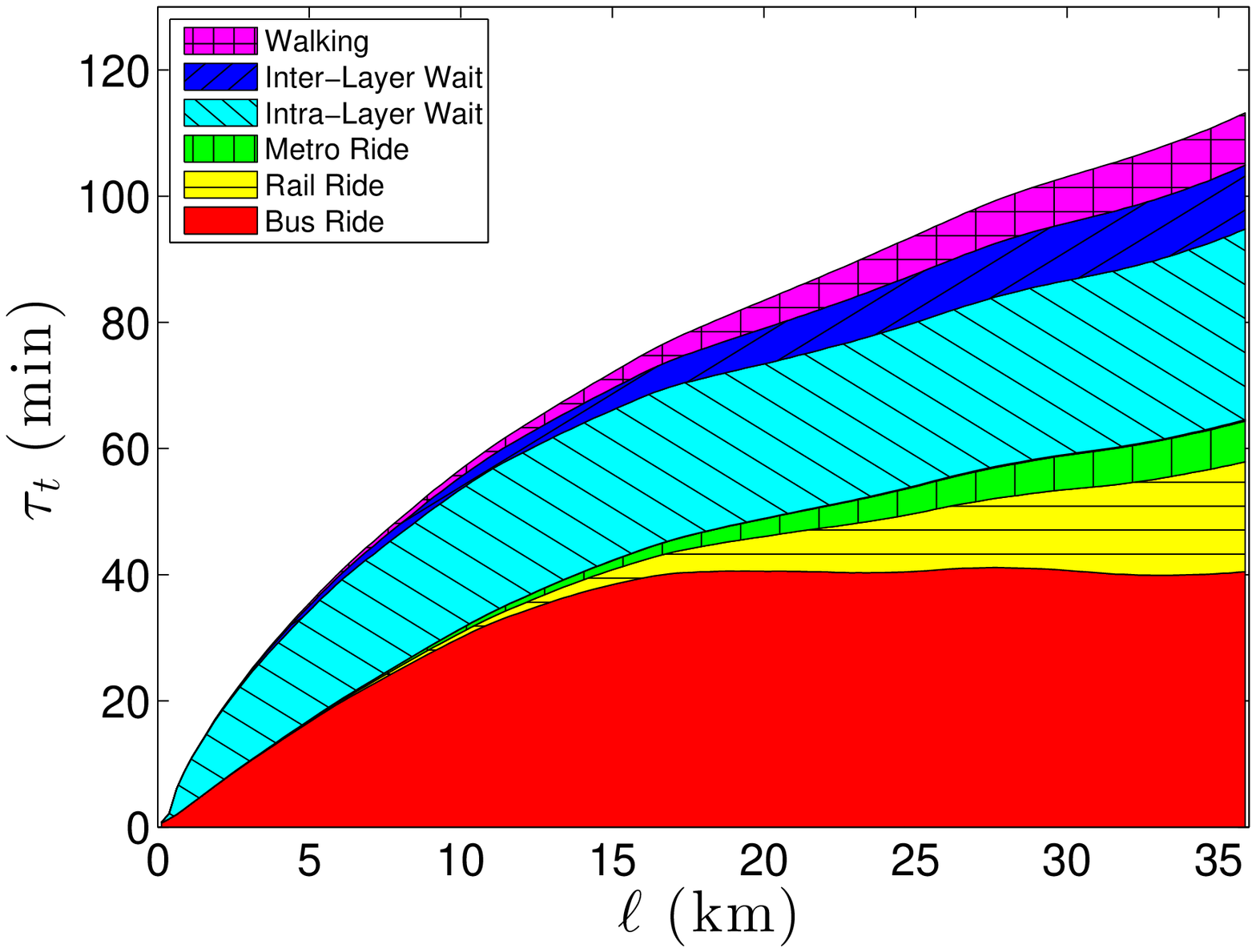} \\
\raisebox{2.3cm}{(c)} \includegraphics[angle=0, width=0.45\textwidth]{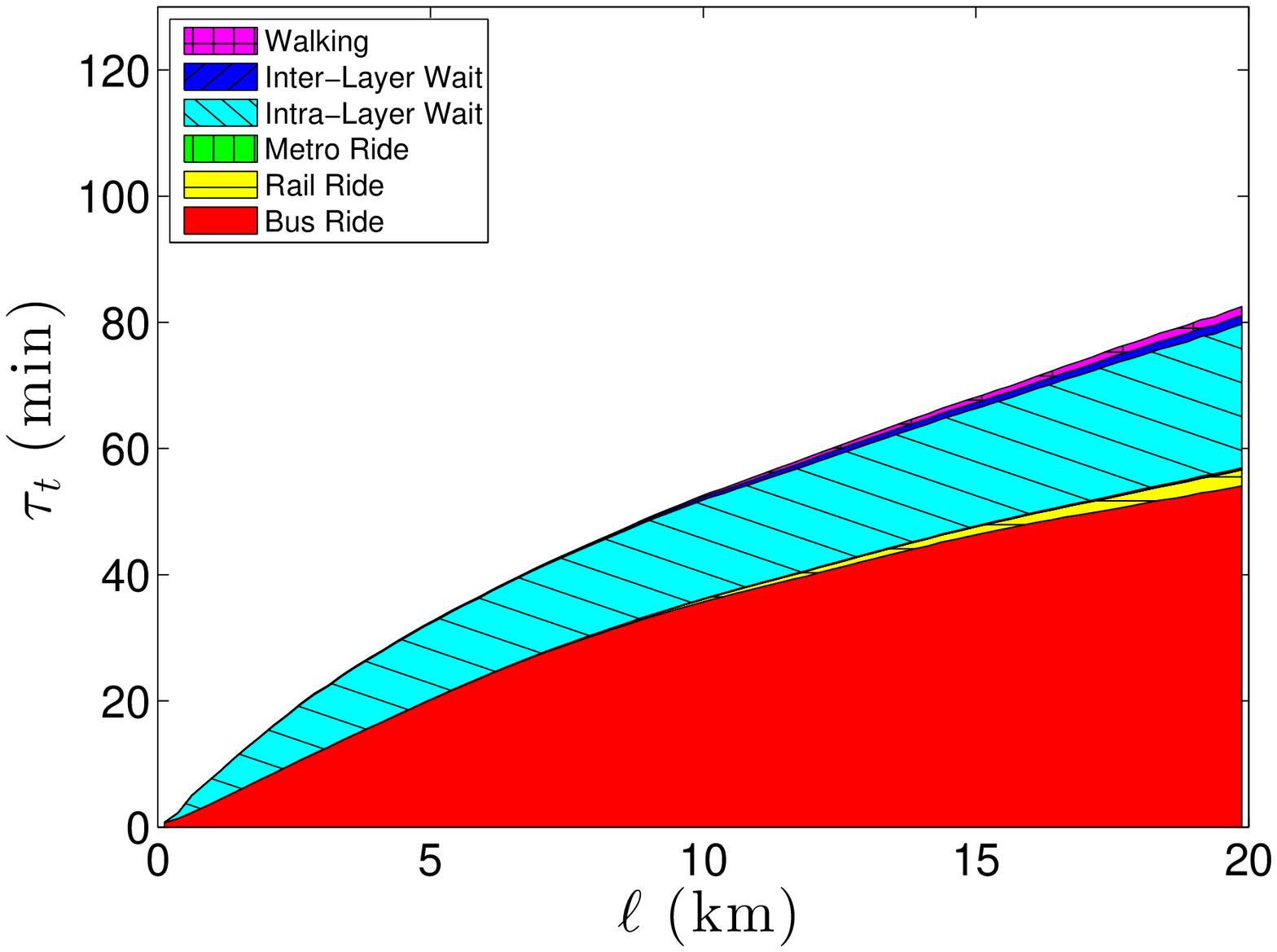} &
\raisebox{2.3cm}{(d)} \includegraphics[angle=0, width=0.45\textwidth]{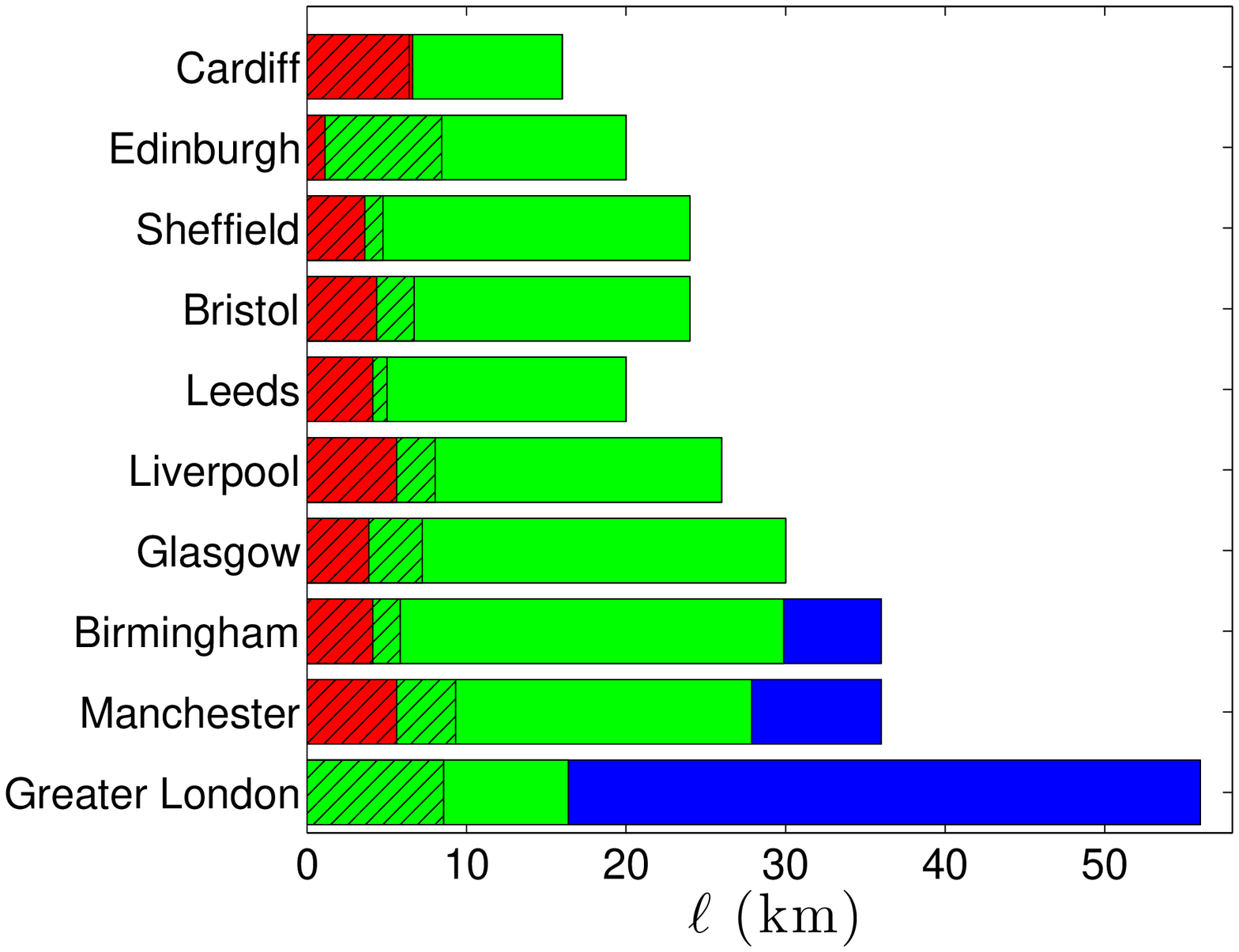} \\
\end{tabular}
\end{center}
\caption{The Anatomy of the transportation networks in (a) London, (b) Manchester, (c) Edinburgh. The average structure of the total travel time changes with the trip length $\ell$ and from a city to another. In all cities but London, waiting times are longer than riding times for short trips. These waiting times are mostly intra-layer waiting times due to bus-bus  interchanges. If the city network is particularly multi-modal, inter-layer waiting times and walking times start playing a significant role for longer values of $\ell$, but only in London do they exceed intra-layer waiting times. Similarly, in all cities but London, the largest fraction of riding times is spent in the bus layer for all distances. In the greater London, for trips longer than $20$kms, most of the riding times are spent in the metro or the rail, demonstrating the importance of these modes in the public transportation sytem of this area. (d) We condense in a single plot some information of the anatomy curves for all cities.  We represent with different colors different regimes versus the trip length: (i) Red: When the trips  are mostly done on the bus layer and display waiting times larger than riding times (a regime not present in London). (ii) Green: when riding times exceed waiting times and most of the distance is covered in the bus layer. (iii) Finally, in blue, when riding times exceed waiting times and most of the distance is covered in the metro and rail layers. The hatched area represents the distances where the overlap $q$ is larger than $50\%$. We observe that when the regime (i) occurs, the overlap is always large and waiting times are then directly added to the minimal paths riding times.}
\end{figure*}


\begin{figure*} 
\begin{center}
\begin{tabular}{cc}
\raisebox{2.3cm}{(a)} \includegraphics[angle=0, width=0.45\textwidth]{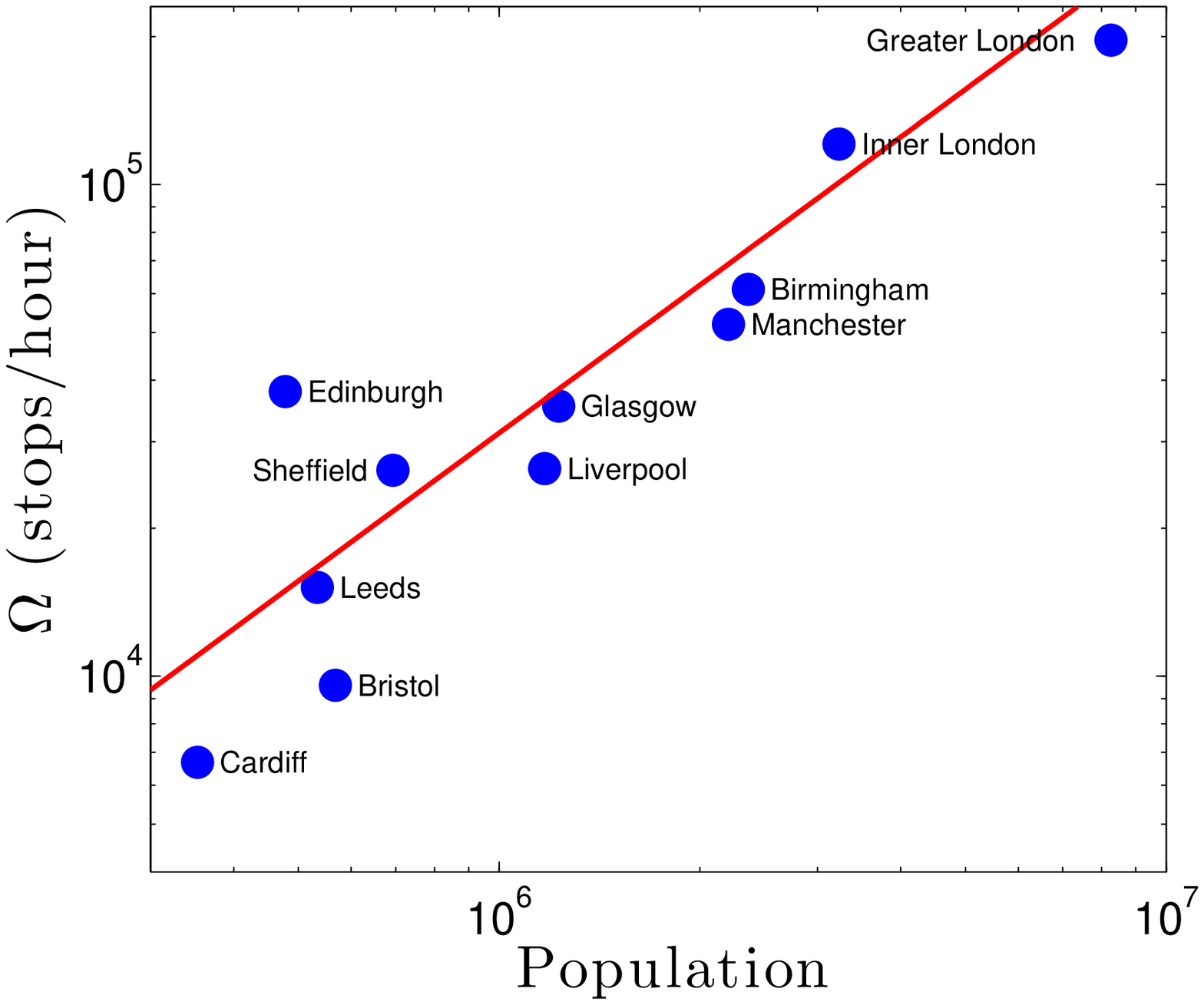}&
\raisebox{2.3cm}{(b)} \includegraphics[angle=0, width=0.45\textwidth]{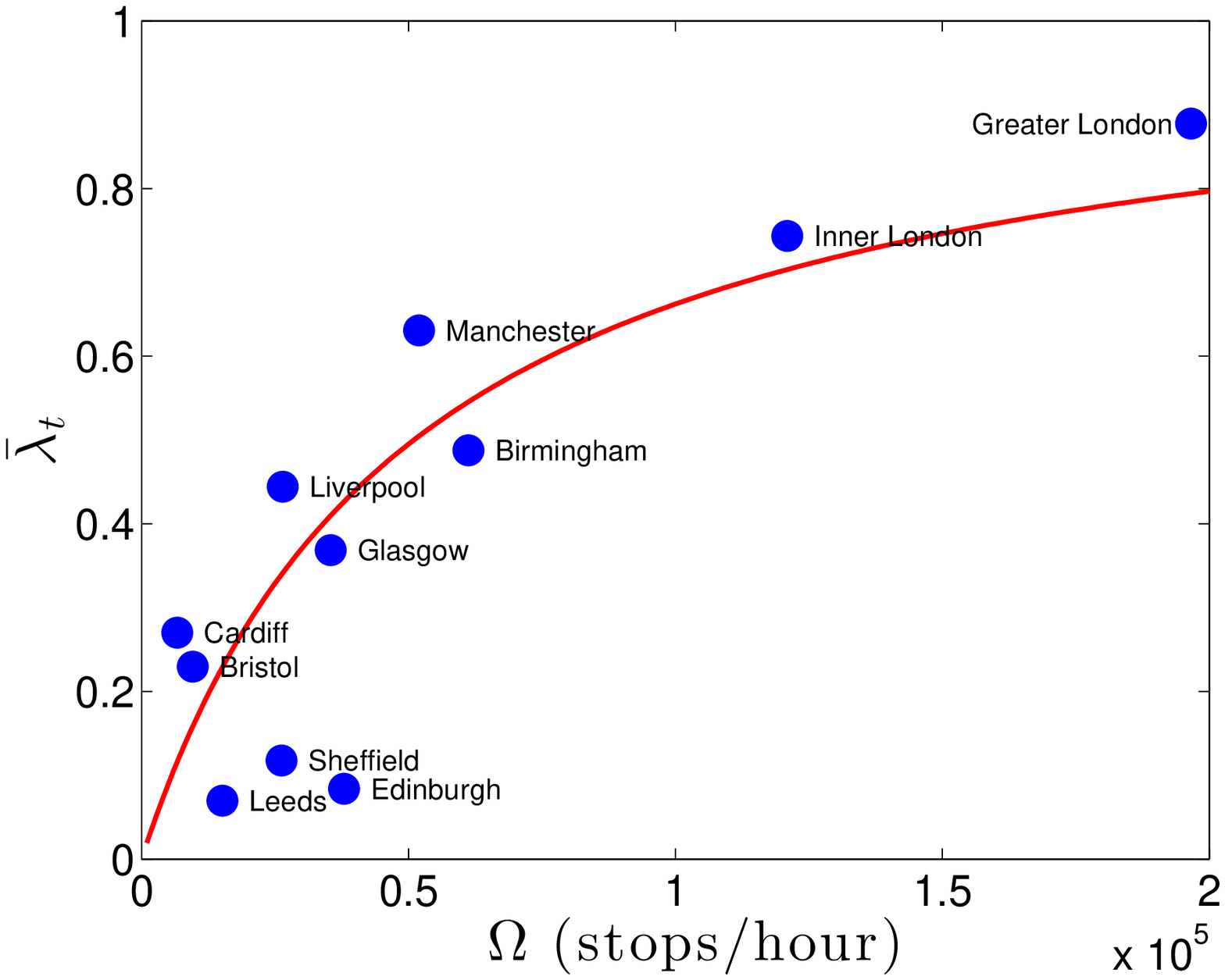}\\
\raisebox{2.3cm}{(c)} \includegraphics[angle=0, width=0.45\textwidth]{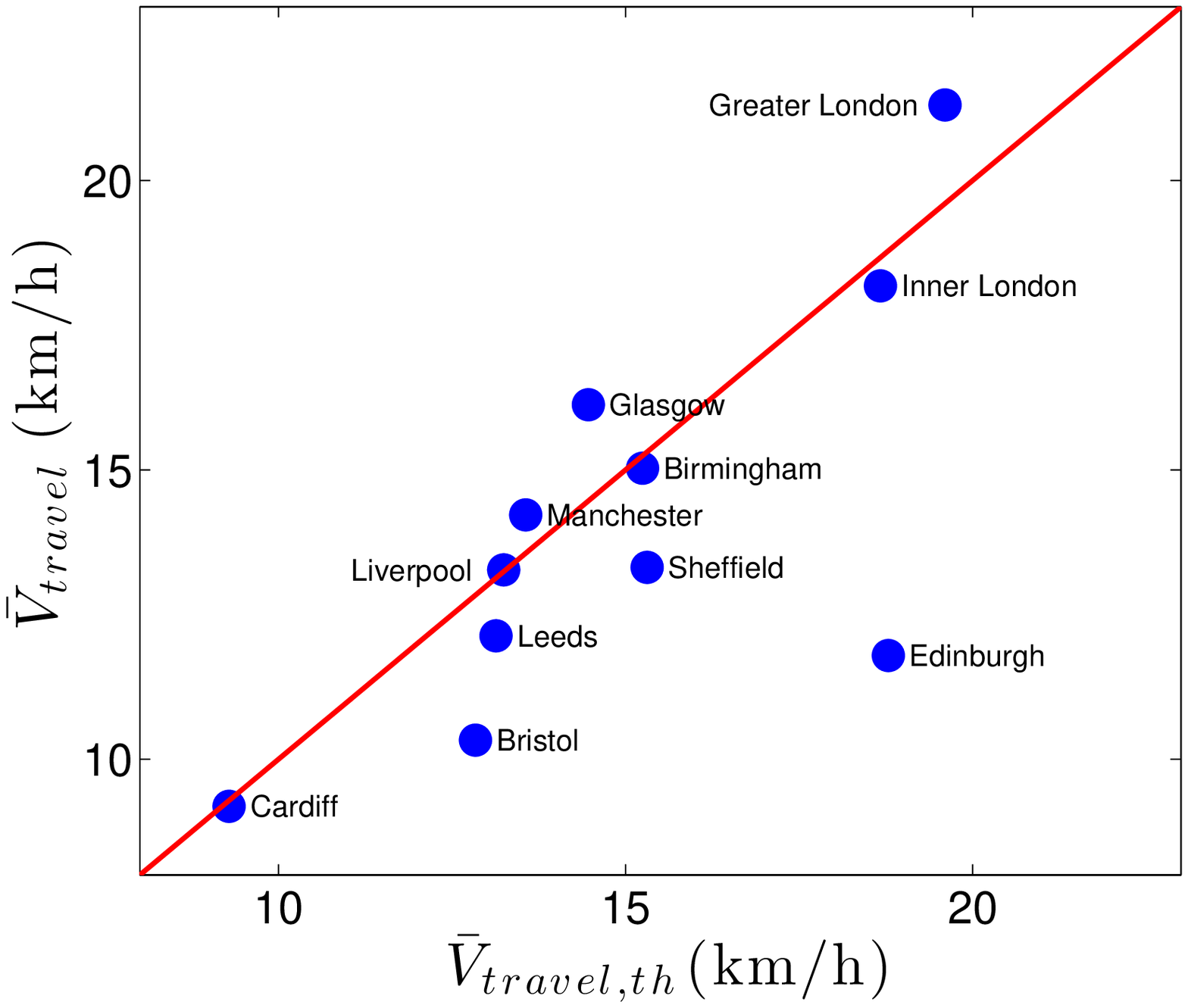} &
\raisebox{2.3cm}{(d)} \includegraphics[angle=0, width=0.45\textwidth]{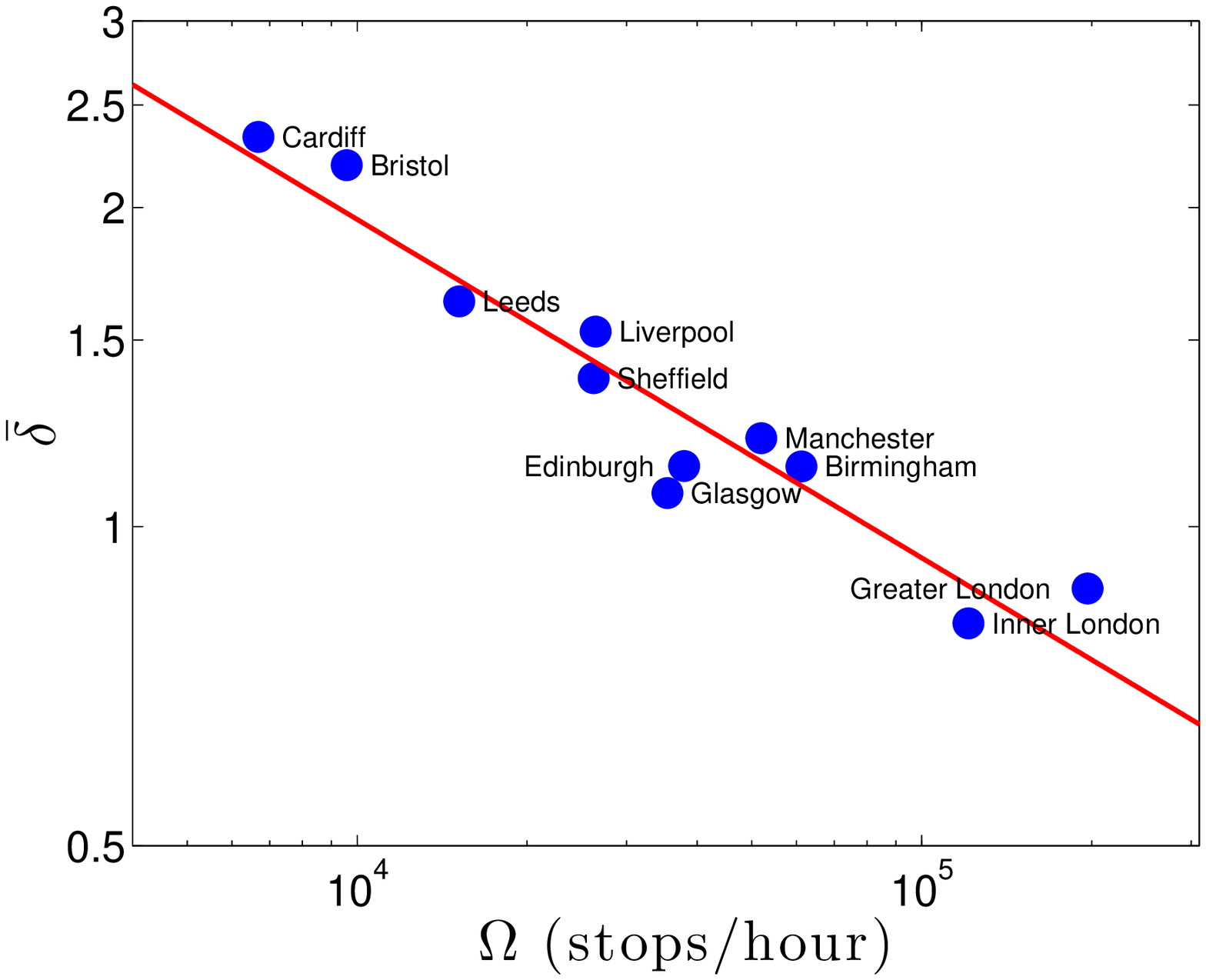} \\
\end{tabular}
\end{center}
\caption{(a) The total number of stop events $\Omega$ grows proportionally with the urban area populations $P$. The power law fit $\Omega\propto P^\phi$ gives $\phi \approx {1.0\pm0.3}$ ($R^2=0.88$). (b) The time-respecting paths interdependency grows as Eq.~\ref{eq:lamb} with $a_{\bar\lambda}=1.65 10^{-5}\ \frac{hour}{stops}$ ($R^2=0.80$), consistent with the hypothesis that the number of possible alternative to exclusive bus layer path is proportional to $\Omega$. (c) The average travel velocity is consistent with Eq. (\ref{vtravel}) characterized by the parameter $k=0.80$ ($R^2=0.87$). (d) The synchronization inefficiency $\delta$  decreases $\Omega$ as a power law $\delta \propto \Omega^{-\mu}$, where $\mu \approx  0.3\pm0.1$ ($R^2=0.91$): time-respecting travel times in larger cities with larger $\Omega$ are closer to the infrastructural limit of minimal travel times.
}
\end{figure*}
\clearpage
\newpage

\section{Supplementary Information}

\subsection{Data Filtering and Elaboration}

Not all Stop Points are actually used, so only those that were present in the timetables are considered active and have been taken into account. Stop points are then organized into Stop Areas representing facilities (Airports, Bus/Metro/Coach/Railway Stations) or possible interchange points. The definition of these Stop areas has been taken as a basis for defining a multi-layer network form the timetable data. 
A further process of data cleaning and aggregation has been performed to have a consistent definition of inter-modal exchange points.

Inconsistent stop times have been corrected by temporal interpolation whenever possible. In the other cases, the stop has been excluded from the dataset. In particular:
i) Many inconsistencies were found in bus stop times: they have been considered wrong whenever two following stops happened more than 2 hours one from another (this applies also the case of decreasing times)
ii) In the original rail timetables, many stop time were erroneously recorded as  `0000': temporal interpolation solved this problem almost entirely.

Inconsistent NaPTAN Stop Areas have been corrected, using as a reference parameter a Walking Distance $wd$ = 500m. The procedure follow those steps:
\begin{itemize}
\item[i)] the Center of an Area, identified by Latitude and Longitude, are corrected with the Stop Points center of mass if before the corrections not all Points were within a distance $wd$ from the Center and after they were;
\item[ii)] Points where Air, Ferry, Rail, Metro and Coach stops happen are always kept in the Area, Bus stops Points further than $wd$ from the Center are removed;
\item[iii)] Areas containing only Bus stops Points are corrected by removing the further stops (and recalculating the center of mass) until they become contained in a circular area of radius $wd/2$ (thus a maximal distance between two points of $wd$);
\item[iv)] Airports Stop Points and Areas are joined together if they share the same first 6 letters in atcocode;
\item[v)] All Air Stop Points are ``promoted'' to Areas;
\item[vi)] The Heathrow Airport stop Area is reconstructed ``by hand'' as the Stop Area was incorrectly defined;
\item[vii)] After imposing a hierarchy [$A>F>R>M>C>B$], all Areas include other Areas and non-bus stops Points of lower rank within a distance $wd$ from its Center (distance between Areas is defined as the distance between their center);
\item[viii)] All remaining non-bus stops Points are ``promoted'' to Areas;
\item[ix)] Rail ,Metro and Ferry Areas of same rank are merged if their distance were under $wd$ (Rail) or $wd/2$ (Ferry,Metro). The choice of $wd/2$ is to avoid joining together following Stops of London Tub and Ferries lines;
\item[x)] All Areas can absorb a Bus stop Point if it is within a distance $wd/2$. In case of conflict, the Point is assigned to the closest;
\item[xi)] Areas containing only one Point are ``declassed'' to Points;
\item[xii)] A stop Point can absorb lower rank Areas/Points if it is within a distance $wd/2$ and become an Area (C and B Points cannot absorb in this step);
\item[xiii)] To each Area is assigned a representing Point, chosen at random between those with higher rank.
\end{itemize}

The so-defined Areas become the inter-layer point for the Multi-layer network. The distance assigned as a inter-layer weight is then computed as the average distance between all Points of the first layer and all Points of the second layer.

{\bf
A copy of this dataset is publicly available at http://www.quanturb.com/data.html
}

\subsection{Cities}

In this paper, we identify as a city a circular area roughly
containing the borders of the associated urban area. Each circle is
defined by the latitude and the longitude of its center and by its
radius (see table below). The value of the population for the relative
Morphological Urban Areas, obtained from wikipedia.org
[http://en.wikipedia.org/wiki/List\_of\_metropolitan\_areas\_in\_the\_United\_Kingdom (Date of access:06/05/2014)]
is associated to each of these areas. For London, two different areas
have been studied. The first one is larger and corresponds to the whole Greater London, while a second, smaller, only the
group of inner boroughs commonly named Inner London
[http://en.wikipedia.org/wiki/Inner\_London (Date of access:06/05/2014)]. Due to our rough
selection of the urban area surface, the value of population used is to be
considered only an approximation of the real population of the
selected surface.
\\

\begin{tabular}{| l | r | r | r | r |r|}
\hline
City Name & Lat & Lon & Radius (km) & Population &$\Omega$ (trips/hour) \\
\hline
Greater London 	&51.51	&-0.12	&28	&8,265,000	& 196,594\\
Inner London 		&51.51	&-0.12	&14	&3,232,000	& 120,937\\
Manchester		&53.48	&-2.24	&18	&2,207,000	&   51,974\\
Birmingham		&52.48	&-1.89	&18	&2,363,000	&   61,225\\
Glasgow			&55.86	&-4.26	&15	&1.228,000	&   35,446\\
Leeds			&53.80	&-1.55	&10	&534,000		&   15,158\\
Liverpool			&53.40	&-2.98	&13	&1.170,000	&    26,444\\
Bristol			&51.45	&-2.58	&12	&568,000		&      9,583\\
Sheffield			&53.38	&-1.42	&12	&693,000		&    26,243\\
Edinburgh			&55.95	&-3.18	&10	&478,000		&    37,942\\
Cardiff			&51.48	&-3.18	&8 	&353,000		&       6685\\
\hline
\end{tabular}
\\
\\

Not all transportation modes are available in every city. Naturally,
air transport is not playing any role and the water transport is
available only in London, Liverpool and Bristol. Moreover, the mode of
transport associated to the Metro layer may be different from a city
to another. In London two types of transportation networks can be
associated to the M-layer: the Underground and Tram. A circular
line of Subways is available also in Glasgow, while in Manchester,
Birmingham and Sheffield the Metro Layer represents the Tram
network. In the other cities considered here, there are no Metro layer.

\clearpage

\subsection{Detour}
\begin{figure*}[h!]
\centerline{
\includegraphics[width=0.45\linewidth]{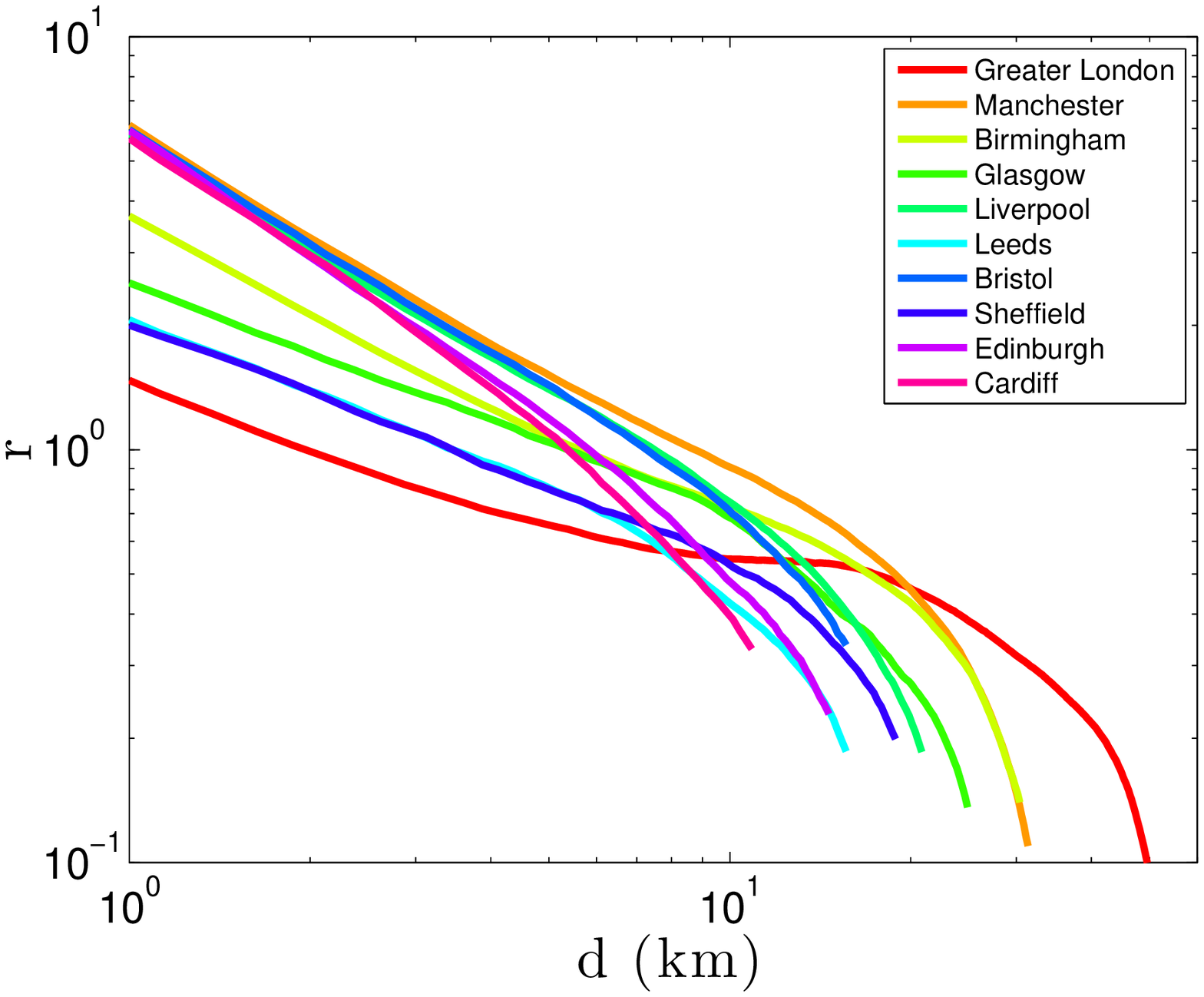}
\qquad
\includegraphics[width=0.45\linewidth]{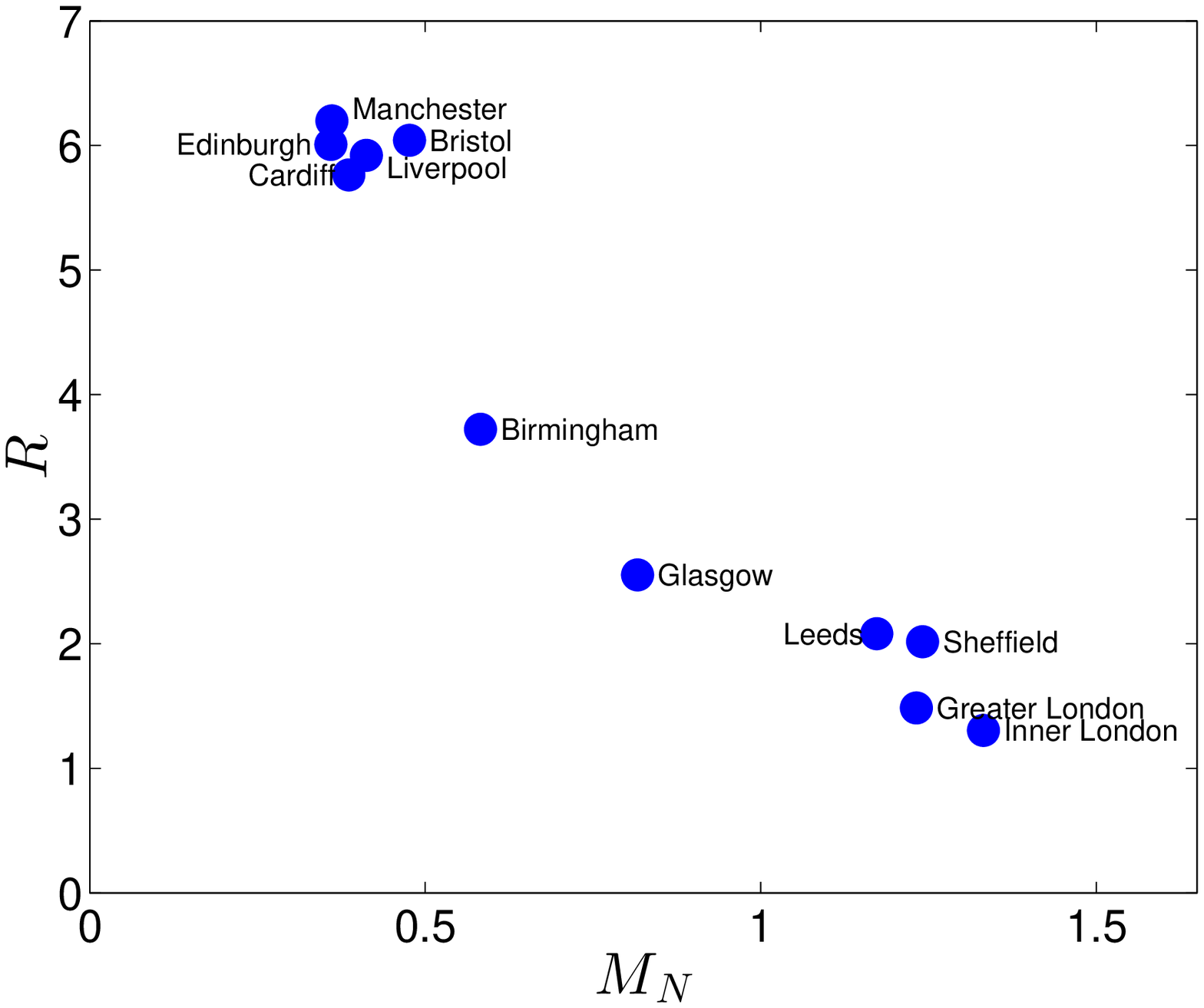}
}
\caption{(Left) Detour Profiles $r(d)$ in different urban
  areas. (Right) The peak value (R) of the detour profiles, decreasing
  with the average cyclomatic number $M_N$.}
\label{SI_det}
\end{figure*}

In both time-respecting paths (figure \ref{SI_det} Left) and
minimal paths, the detour $r(d) = \ell(d)/d$, where $\ell$ is the
effective path's length and $d$ the euclidean distance between origin
and destination, is a decreasing function of $d$. In our
analysis, we exclude trips where origin and destination are less than
1 km apart, and the quantity $R=\max_{d>1} r(d)$ corresponds in our case
to the detour for the shortest considered path $r(1km)$. In figure
\ref{SI_det} Right, we see that $R$ decreases when the average
cyclomatic number $M_N = (E-N-1)/N$ grows, where $E$ is the number of
directed edges and $N$ the number of nodes of the network. This
suggests that the availability of more alternatives, characterized by
a larger number of cycles per node in the network, makes straighter
trajectories possible.

\clearpage

\subsection{Connection times in multi-modal trajectories}

\begin{figure*}[h]
\centerline{\includegraphics[width=0.45\linewidth]{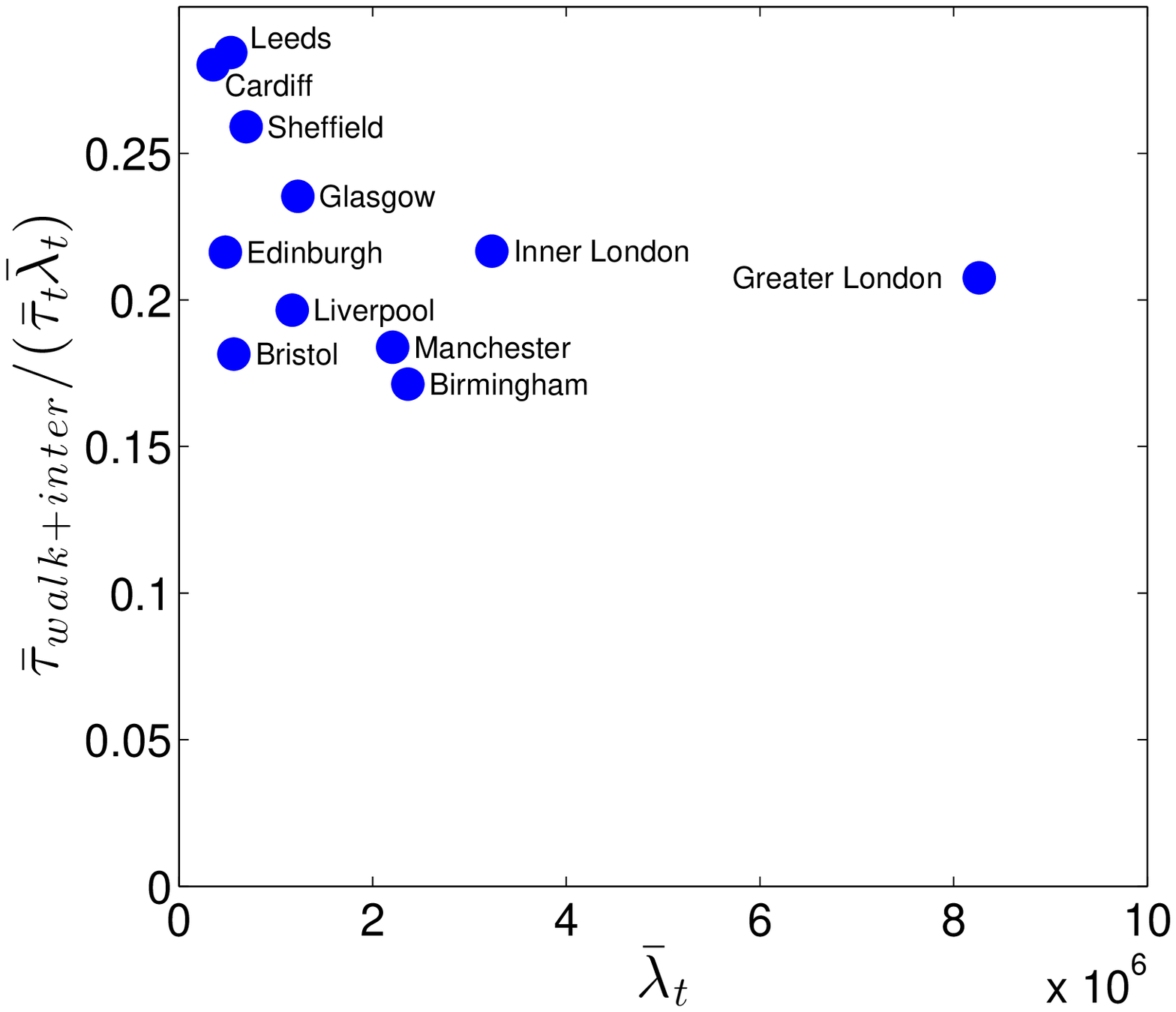}}
\caption{{\bf The average inter-layer connection time for a city, defined as the sum of the walking time and the inter-layer waiting time $\bar \tau_{walk+inter}$ over the total traveltime $\bar \tau$ clearly depends upon the fraction $\bar\lambda_t$ of trajectories that have inter-layer connections. Dividing by $\bar\lambda_t$, we estimate the average fraction of inter-layer connection time, restricted to the interdependent trajectories. We notice that, for cities of different size and offer for transport service, this value is relatively stable and all values can be found in the interval $23\%\pm6\%$.
}}
\end{figure*}

\subsection{Time-respecting Paths Travel Speeds}

\begin{figure*}[h!]
\centerline{\includegraphics[width=0.45\linewidth]{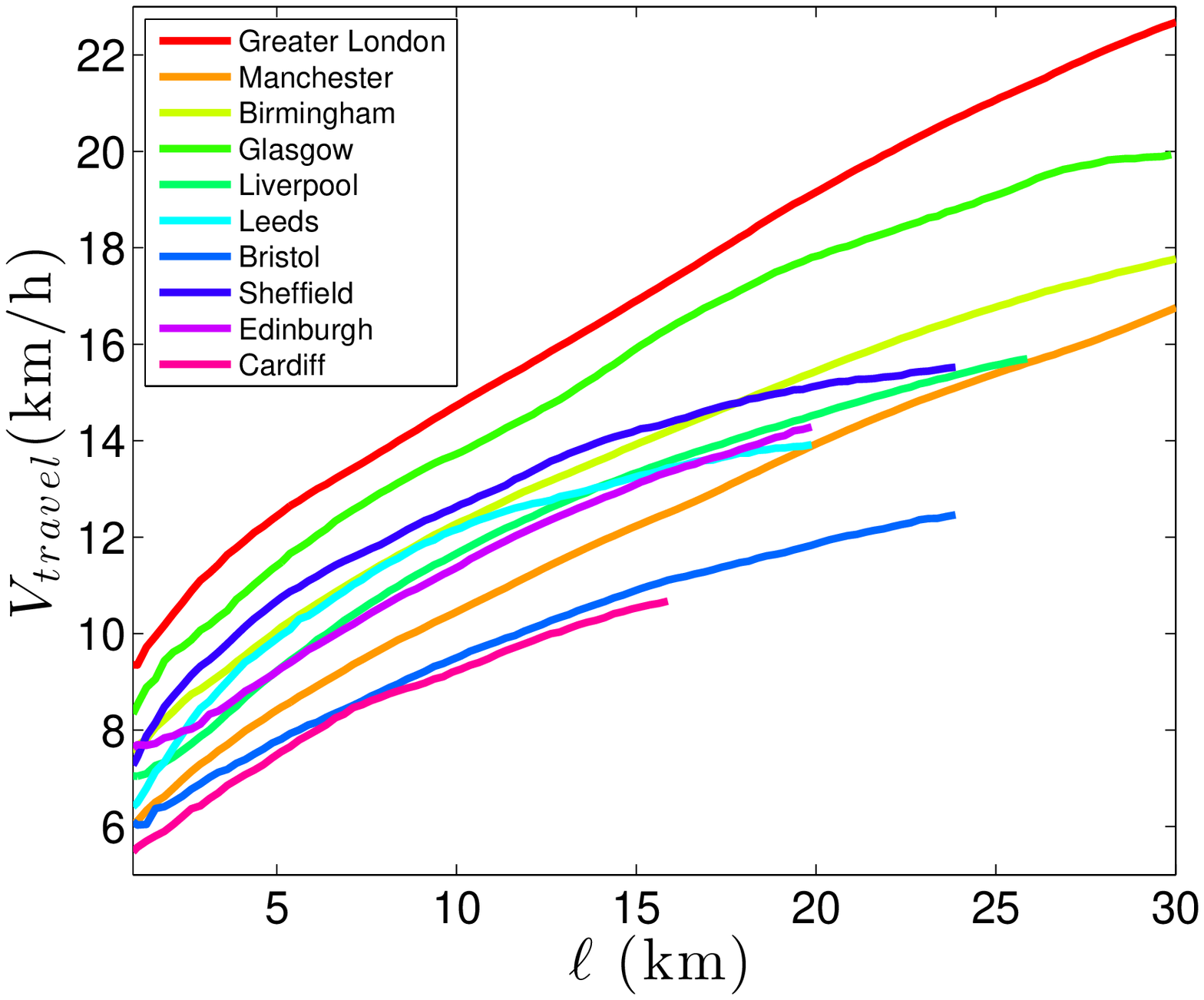}}
\caption{The travel speed $V_{travel} = \ell/\tau_t$ grows with the trip's length $\ell$ and seem not to reach any saturation value within the urban areas' radios. In the main paper, we show that the average value $\bar V_{travel}$ is linked to the interdependency $\lambda$ and, as a consequence, to the average number of stop events per hour $\Omega$ or to the urban area population.}
\end{figure*}

\clearpage

\subsection{Average speeds and frequencies}
Let  $e_k^\alpha = (i,j)$ be a directed edge from a vertex $i$ to a second vertex $j$, both belonging to layer $\alpha$. The edge is identified by the index $k = 1\dots E_\alpha$. Each edge is characterized a the speed $V(e^\alpha_k)$ and the frequency $f(e^\alpha_k)=C(e^\alpha_k)/\Delta t$, where $C(e^\alpha_k)$ is the number of departure events through the edge $e^\alpha_k$ in the time interval $\Delta t$. As our study is focussed on the mobility starting at 8:00 of a working Monday, we chose as extremes of the time interval $t_{end}=$ 24:00 $t_0=$ 08:00 of Monday, and thus $\Delta t = 16$h.

For every city, we may define the average speed of a layer $\alpha$ as the average over all edges's speed for that layer:
\begin{equation}
\bar V_\alpha = \sum_{k=1}^{E_\alpha} V(e^\alpha_k)/E_\alpha
\end{equation}
and, similarly, the average frequency is:
\begin{equation}
\bar f_\alpha = \sum_{k=1}^{E_\alpha} f(e^\alpha_k)/E_\alpha
\end{equation}

\begin{figure*}[h!]
\centerline{
\includegraphics[width=0.9\linewidth]{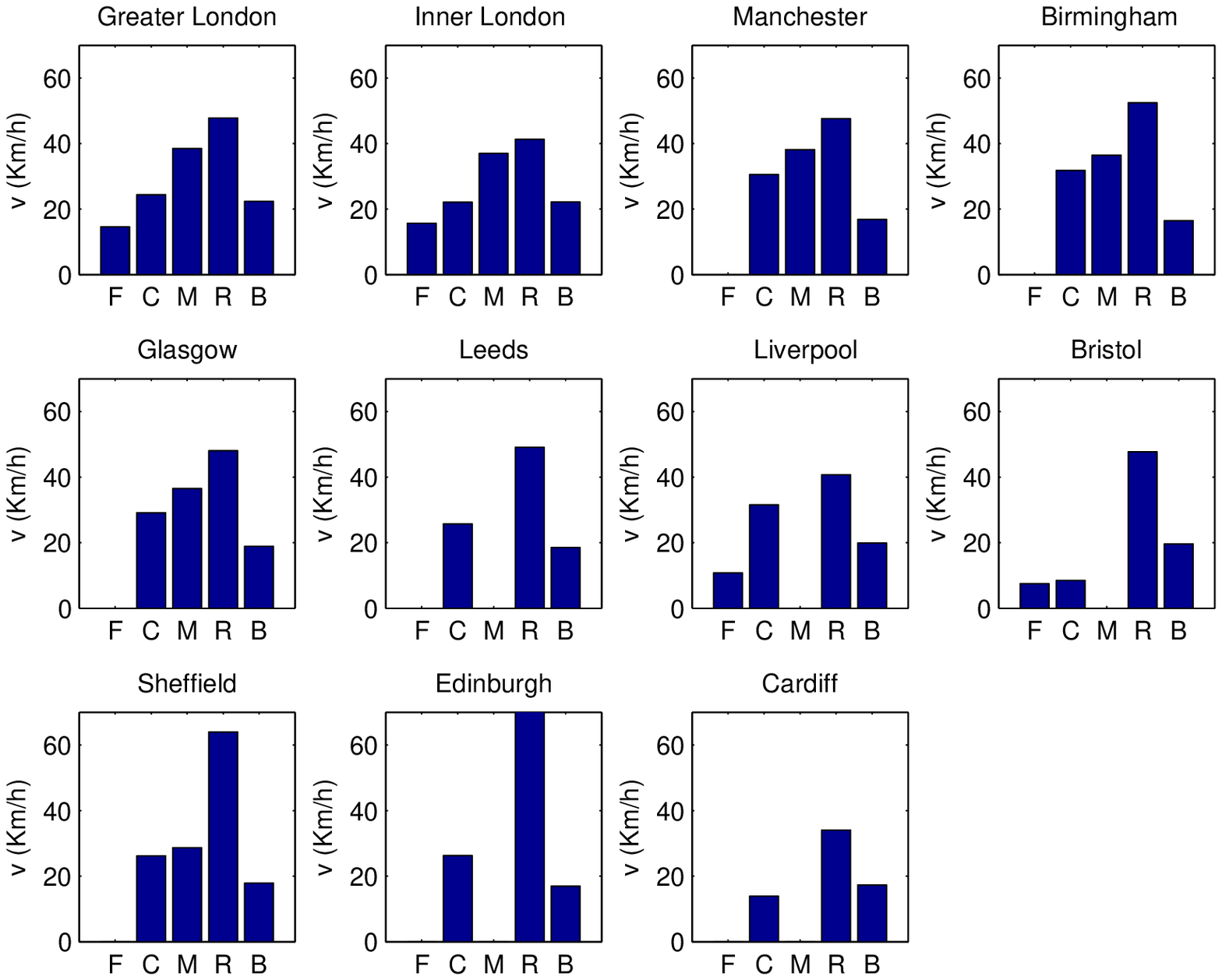}
}
\label{SIspeed}
\caption{Average layer edges' speed $V_\alpha$ in the different Public Transport Networks: F=Ferry, C=Coach, M-Metro, R=Rail, B=Bus. Note that the Ferry and Metro layer are not available in all cities.}
\end{figure*}

\begin{figure*}[h!]
\centerline{
\includegraphics[width=0.9\linewidth]{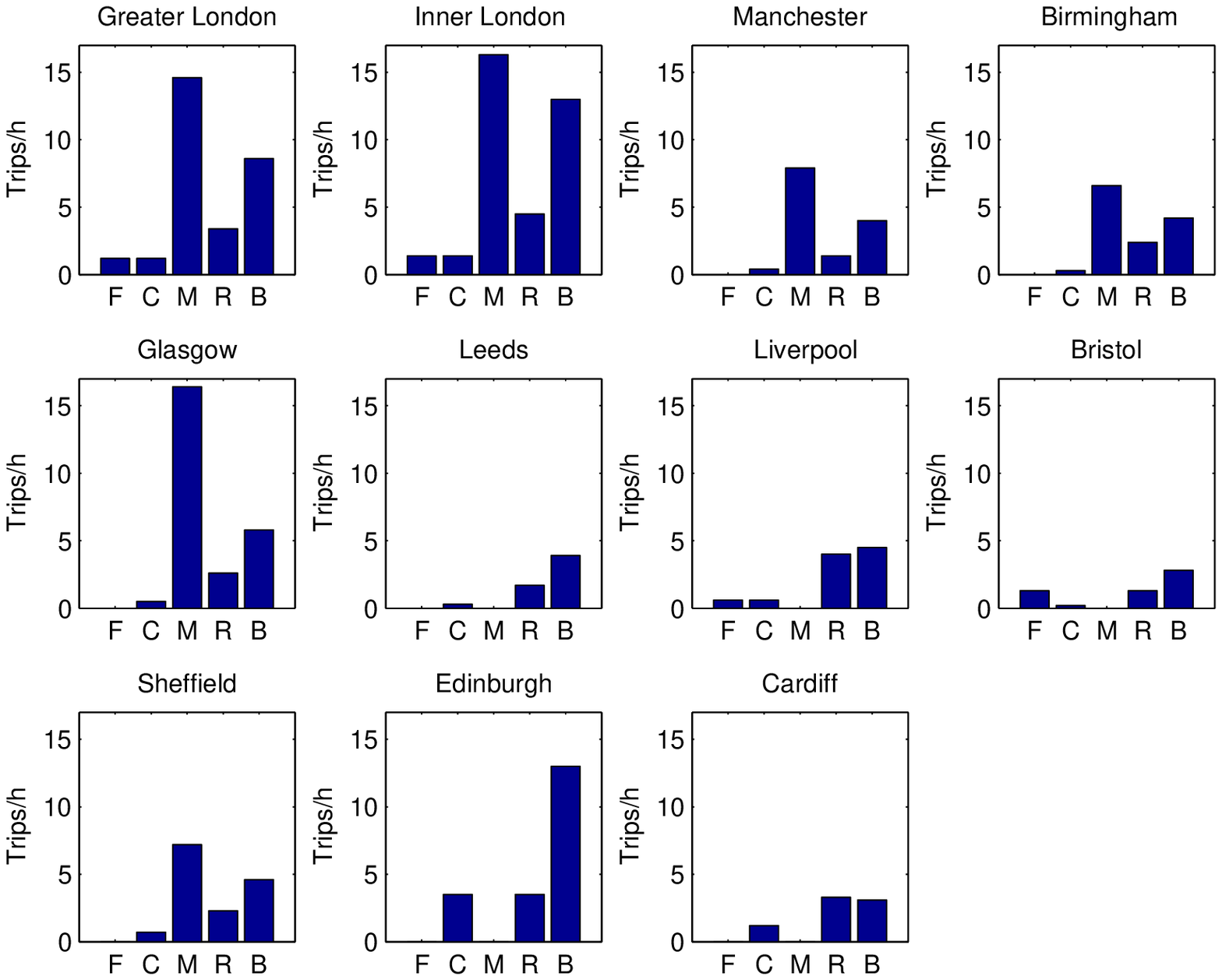}
}
\label{SIfreq}
\caption{Average layer edges' frequencies $f_\alpha$ in the different Public Transport Networks: F=Ferry, C=Coach, M-Metro, R=Rail, B=Bus. Note that the Ferry and Metro layer are not available in all cities.}
\end{figure*}

In figures \ref{SIspeed} and \ref{SIfreq} we see how these quantities differ significantly between the considered cities.


\begin{figure*}
\centerline{
\includegraphics[width=0.45\linewidth]{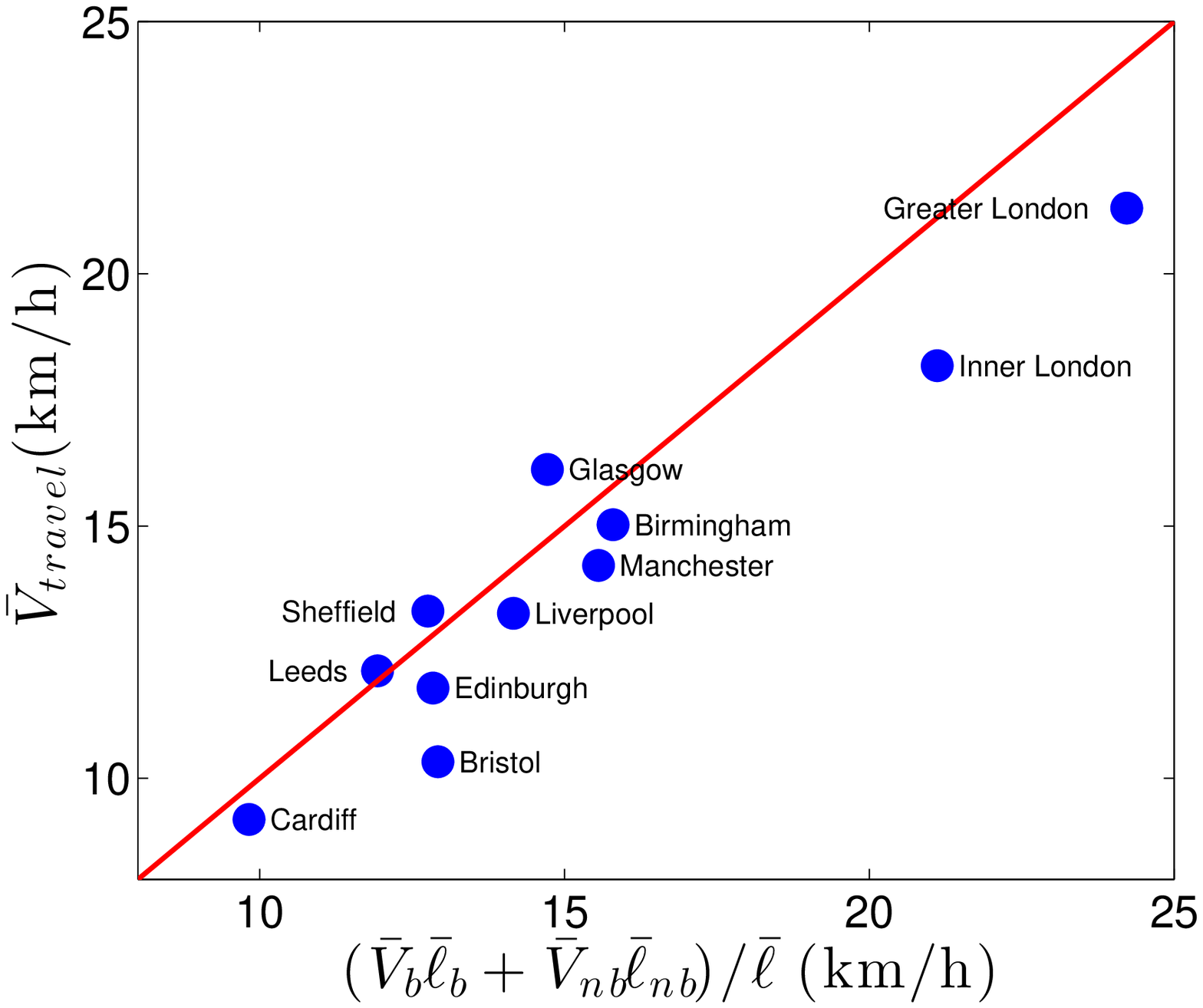}
\qquad
\includegraphics[width=0.45\linewidth]{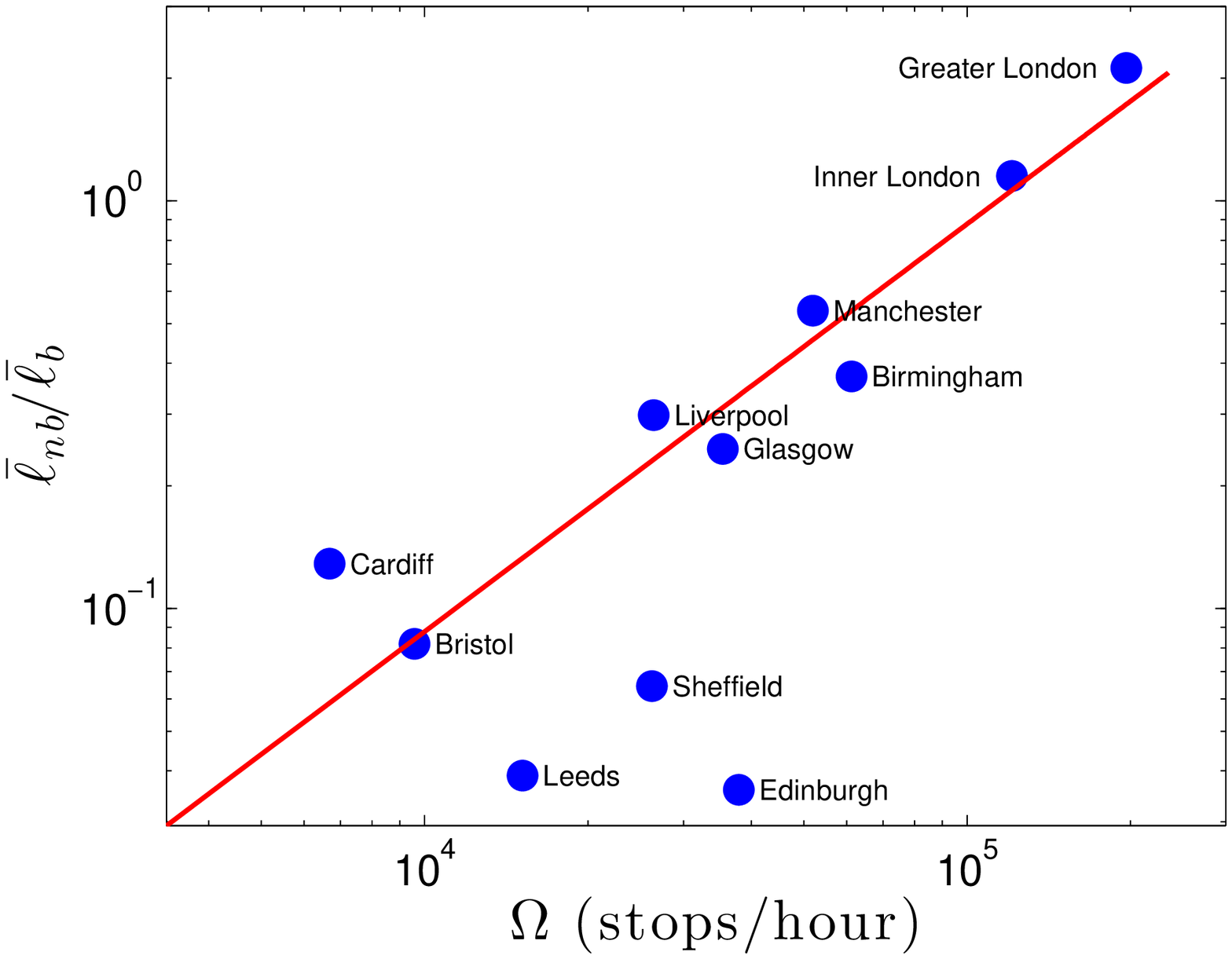}
}
\caption{(Left) The average cruise speeds can be estimated by knowing for each layer: i) the average link's speeds; ii) the fraction of length covered in that layer. (RIght) The ratio $\bar f_{nb}/\bar f_b$ grows with $\Omega$. The solid line is a guide for the eye suggesting a direct proportion.}
\label{SIdeco}
\end{figure*}

In our time-respecting paths, only the Bus, Rail and, when available, Metro layer play a major role. To compare the results of cities where the Metro layer is available with cities where it is not, we introduce the average non-bus speed $\bar V_{nb}$ and frequency $\bar f_{nb}$ defined as:

\begin{equation}
\bar V_{nb} =\frac{\bar \ell_m}{\bar\ell}  V_m +  \frac{\bar\ell_r}{\bar\ell} V_r
\end{equation}
and the average frequency:
\begin{equation}
\bar f_{nb} =\frac{\bar\ell_m}{\bar\ell}  f_m +  \frac{\bar\ell_r}{\bar\ell} f_r
\end{equation}
where$ \frac{\bar\ell_m}{\bar\ell}$ and $\frac{\bar\ell_r}{\bar\ell}$ are, for each city, the average on all trips of the fraction of the total length $\ell$ that is covered on the Metro (m) or Rail (r) layer respectively. As we see in figure \ref{SIdeco}, this simple proportion permits us to reconstruct reasonably well the differences of cruise speed $V_{cruise}=\langle\ell/\tau_ride\rangle$ in different cities.

\subsection{Average interdependency and speed}

Hypothesis: the number of possible alternative to exclusive bus layer path (which is always an available option) is $\propto \Omega$.
Thus, fraction of paths using only the bus layer is
\begin{equation*}
1-\lambda_t = \frac{1}{1+a_\lambda\Omega}
\end{equation*}
and conversely
\begin{equation*}
\lambda_t = \frac{a_\lambda\Omega}{1+a_\lambda\Omega}
\end{equation*}

Hypothesis: once the non-bus layers are involved, there is a proportion between the distance covered in the bus $\ell_b$ ant the distance covered in the non-bus $\ell_nb$
\begin{equation*}
k = \frac{\ell_{nb}}{\ell_b} 
\end{equation*}
With this assumption, $V_{cruise}$ reads

\begin{align*}
V_{cruise} &= V_{b}(1-\lambda_t) +  \frac{V_b\ell_{b}   +V_{nb}\ell_{nb}}{\ell_b+\ell_{nb}}\lambda_t\\
		 &=\frac{ V_{b} +\frac{\frac{V_b}{V_{nb}}+k}{1+k}  V_{nb}a_\lambda\Omega} {1+a_\lambda\Omega}
\end{align*}

\subsection{Contributions to the inefficiency}
\begin{figure*}
\centerline{
\includegraphics[width=0.45\linewidth]{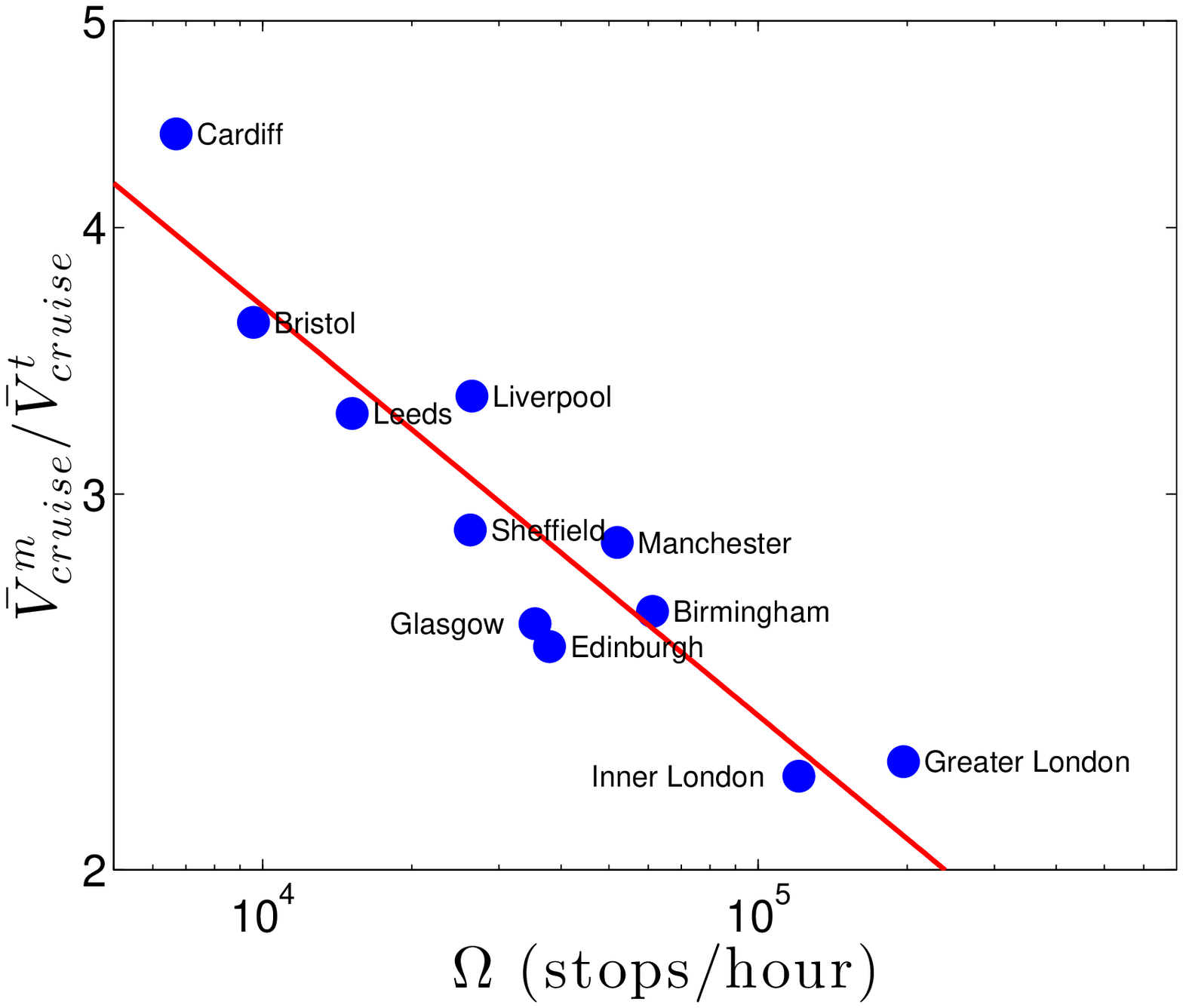} 
\qquad
\includegraphics[width=0.45\linewidth]{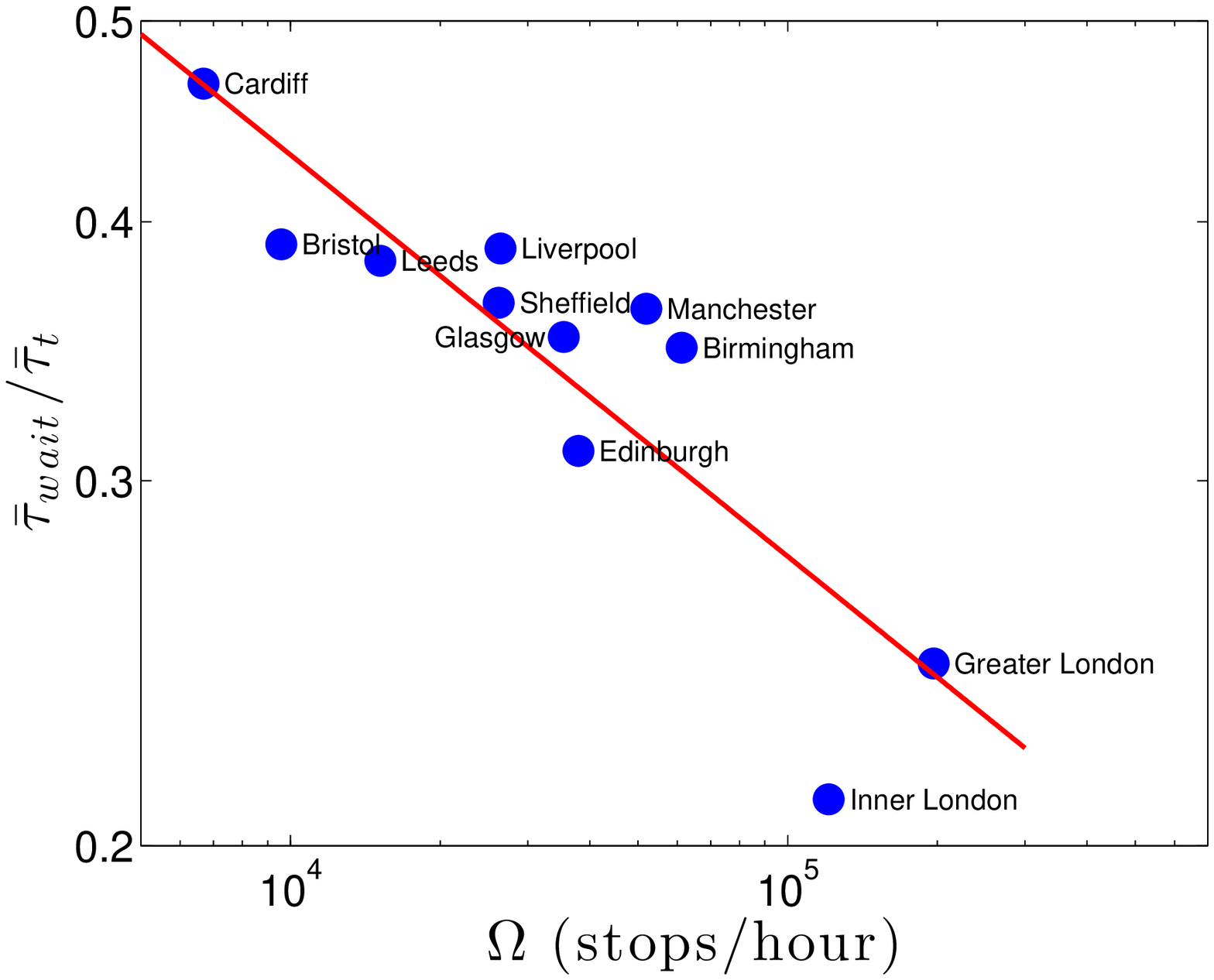} 
}
\caption{(Left) The ratio between the average cruise speed in time-respecting paths $V_{cruise}^t$ and in minimal paths$V_{cruise}^m$ falls with $\Omega$  as $ V_{cruise}^t/V_{cruise}^m \propto \Omega^{-0.19\pm0.06}$ ($R^2=0.87$). (Right) The relative weight of waiting times $\tau_{wait}$ over the total travel times $\tau_t$ for time-respecting paths decreases also as  $\tau_{wait}/\tau_t \propto \Omega^{-0.19\pm0.08}$ ($R^2=0.76$).
}
\label{SIineffCont}
\end{figure*}

In our paper, we show that the average synchronization inefficiency $\delta = \tau_t/\tau_m-1$ decreases when $\Omega$ grows. More specifically, there are two factors contributing to a lower inefficiency: cruise speeds in the time-respecting paths become progressively closer to those of minimal paths (fig. \ref{SIineffCont} Left), and a lower relative contribution of the waiting times (inter- and intra-layer) to the total travel time (fig.  \ref{SIineffCont} Right).

\subsection{Walking time and Multi-modality}

\begin{figure*}
\centerline{
\includegraphics[width=0.45\linewidth]{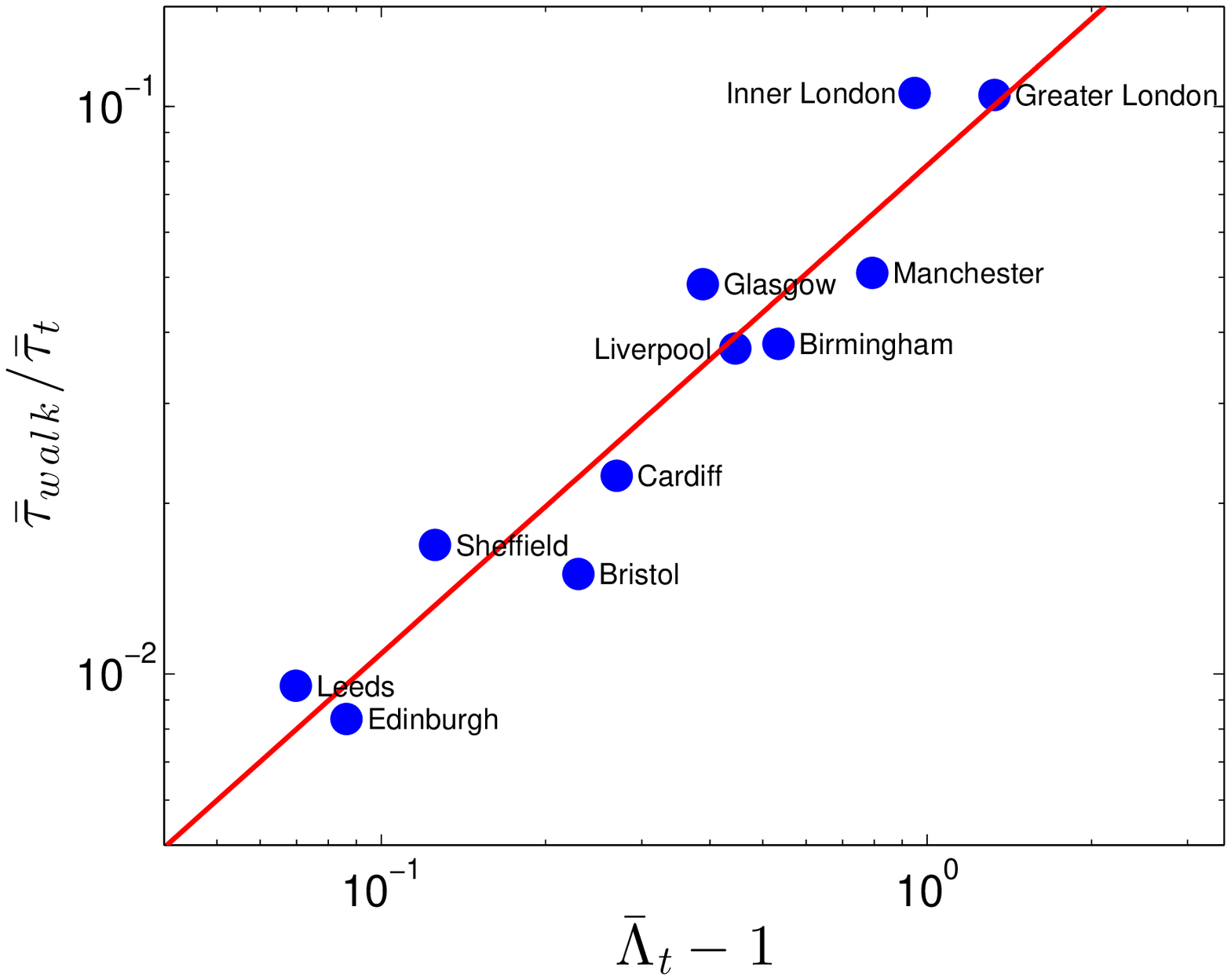} 
}
\caption{As one may expect, the fraction of travel time spent walking grows with the average number $\bar\Lambda_t$ of layers (and thus of connections) involved in the time-respecting paths. The growth is consistent with a direct proportionality: $\tau_{walk}/\tau_t \propto \Lambda^{-0.9\pm0.2}$ ($R^2=0.92$)
}
\end{figure*}

\subsection{Anatomies} 

Here below we complete the overview of the Anatomy of the transport networks described in figure 6 of the Paper.

\begin{figure*}
\begin{center}
\begin{tabular}{cc}
\raisebox{2.3cm}{(a)} \includegraphics[angle=0, width=0.45\textwidth]{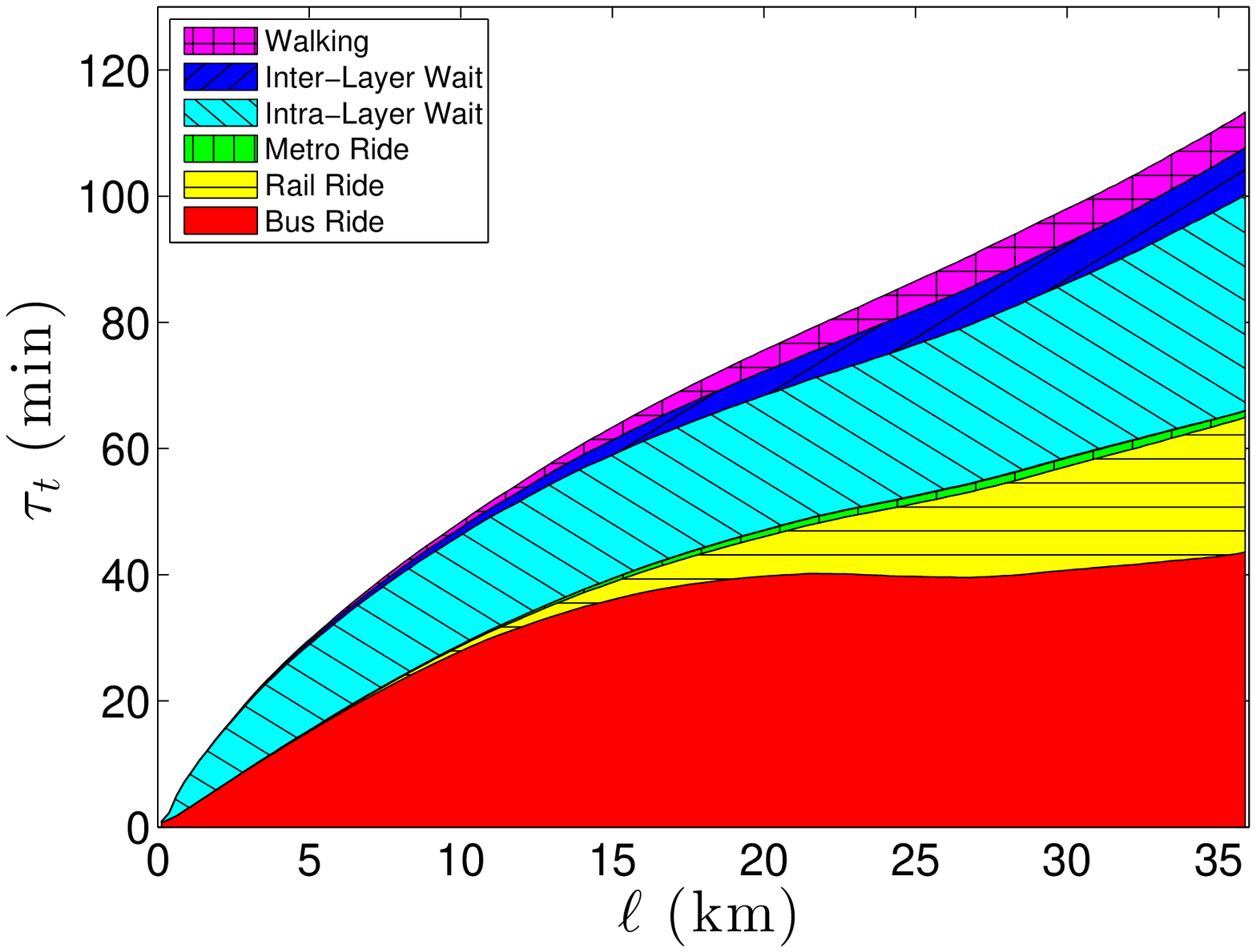}&
\raisebox{2.3cm}{(b)} \includegraphics[angle=0, width=0.45\textwidth]{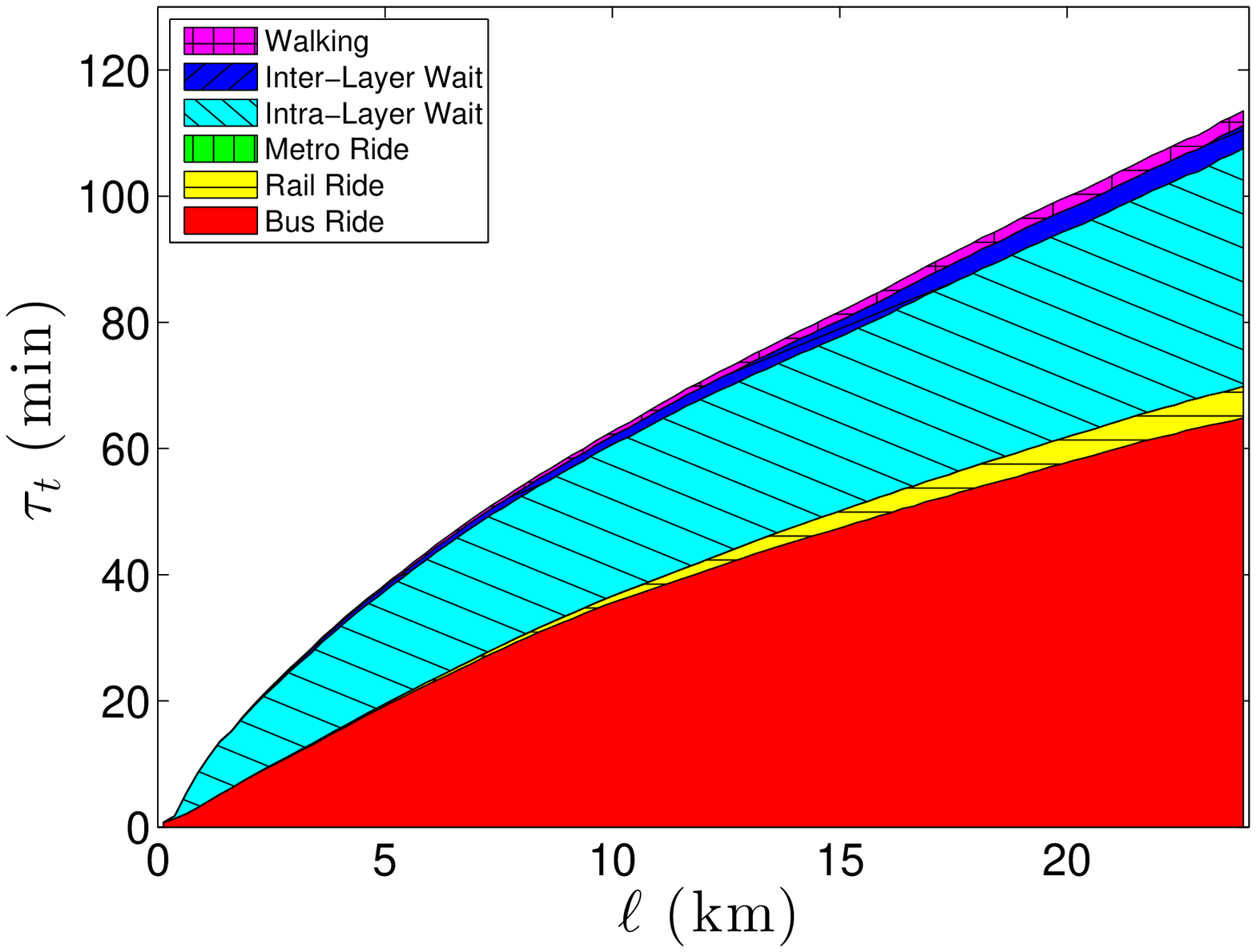} \\
\raisebox{2.3cm}{(c)} \includegraphics[angle=0, width=0.45\textwidth]{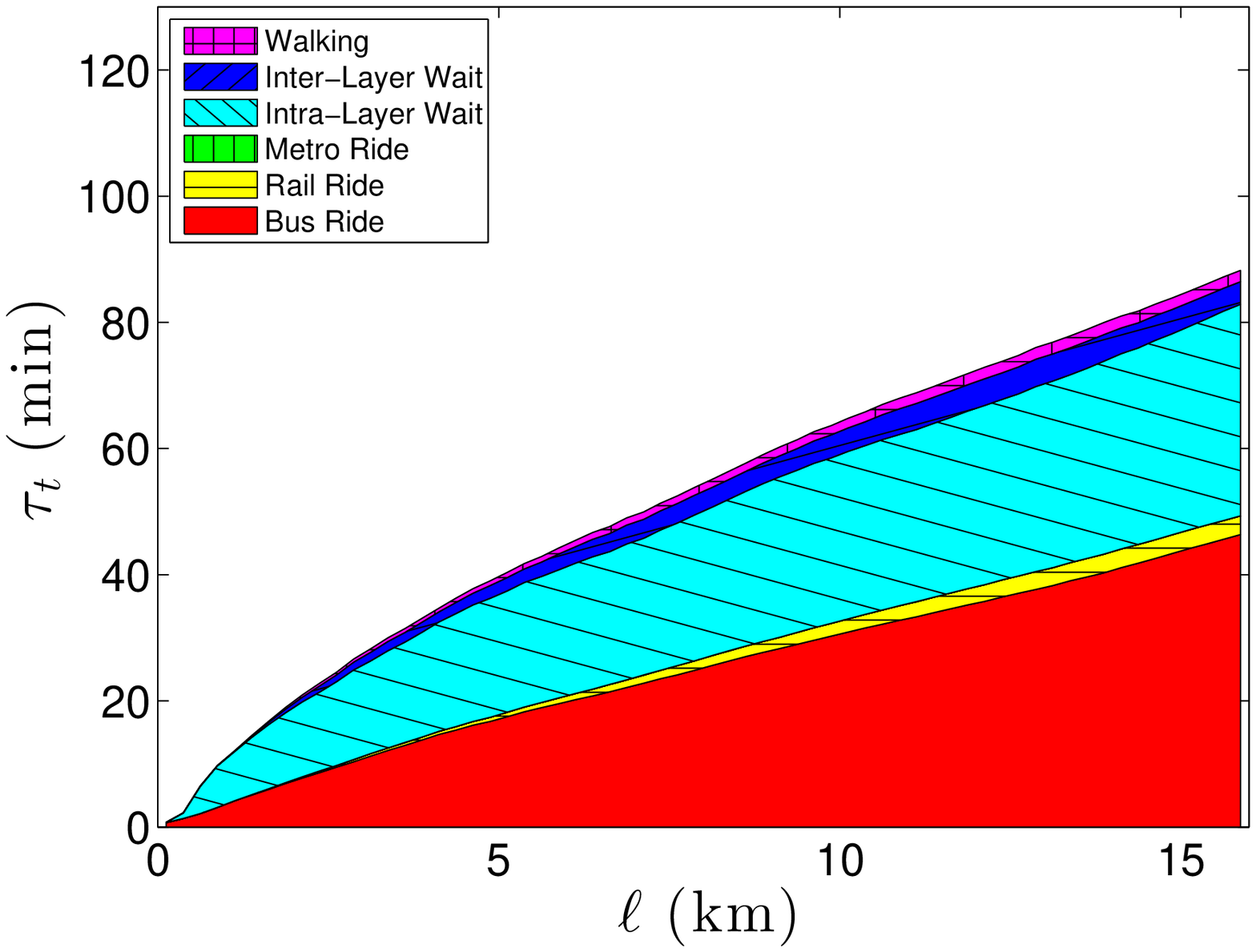} &
\raisebox{2.3cm}{(d)} \includegraphics[angle=0, width=0.45\textwidth]{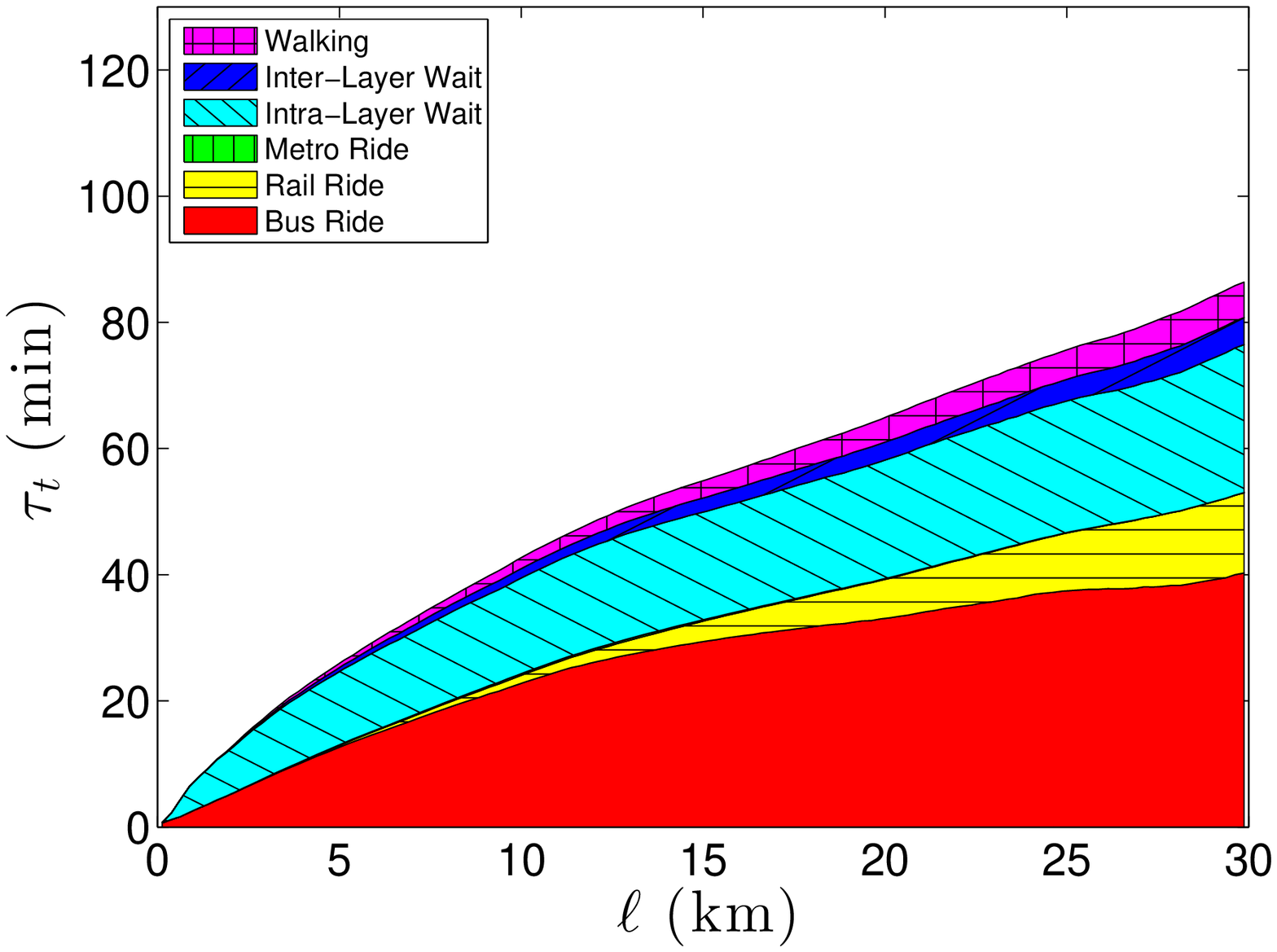} \\
\end{tabular}
\end{center}
\caption{The Anatomy of the transport networks in: (a) Birmingham; (b) Bristol; (c) Cardiff; (d) Glasgow}
\end{figure*}

\begin{figure*}
\begin{center}
\begin{tabular}{cc}
\raisebox{2.3cm}{(a)} \includegraphics[angle=0, width=0.45\textwidth]{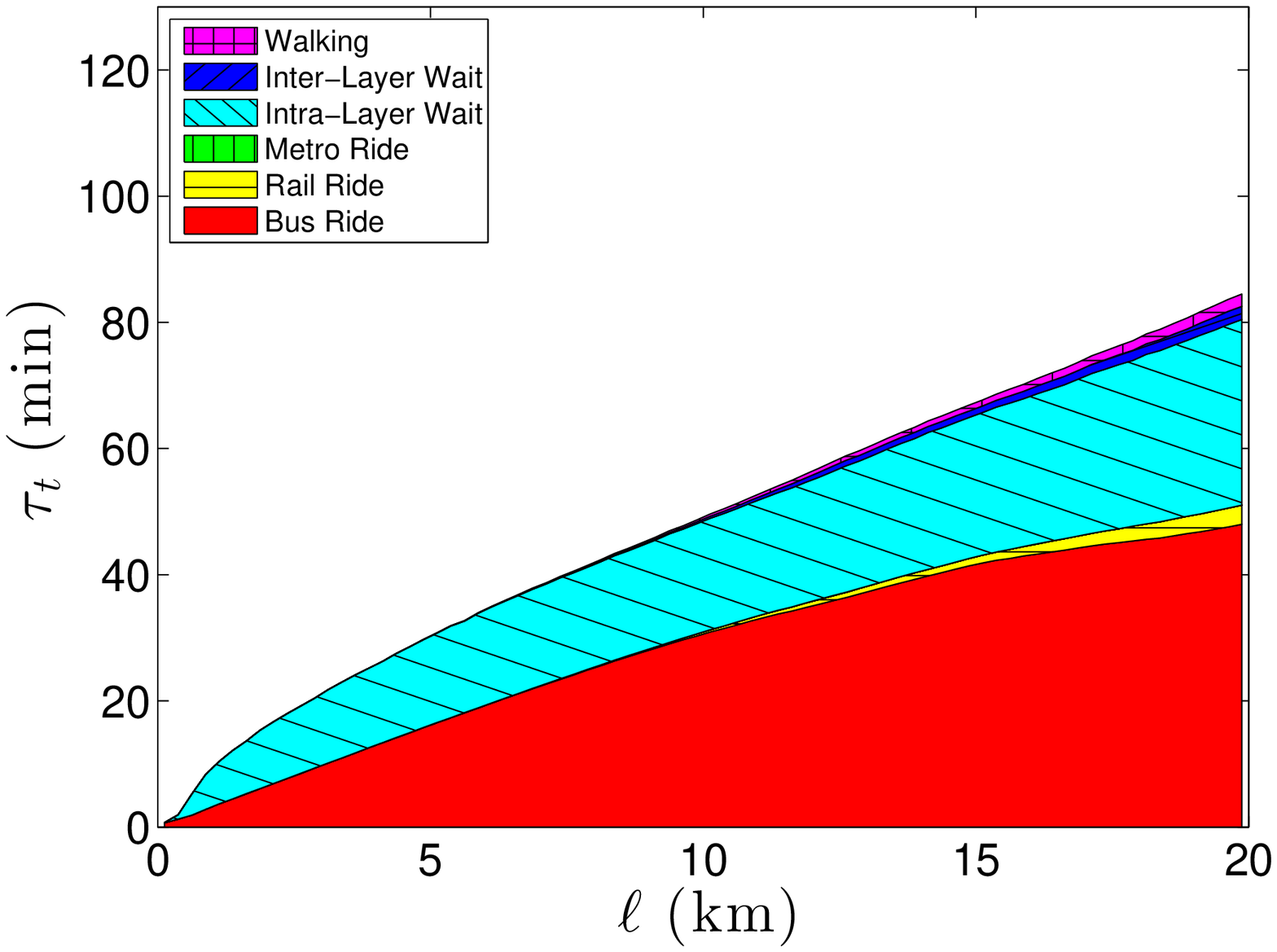}&
\raisebox{2.3cm}{(b)} \includegraphics[angle=0, width=0.45\textwidth]{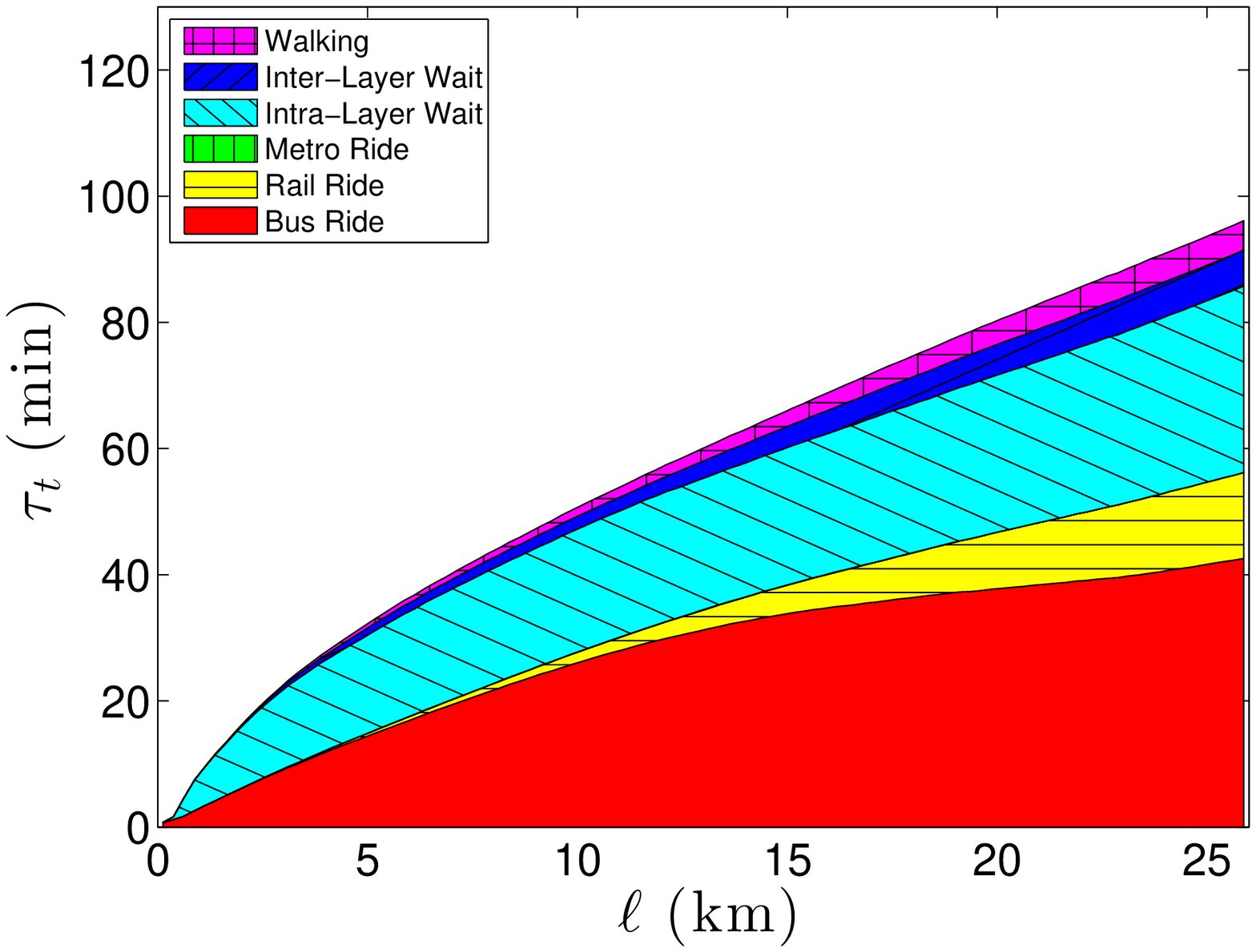} \\
\raisebox{2.3cm}{(c)} \includegraphics[angle=0, width=0.45\textwidth]{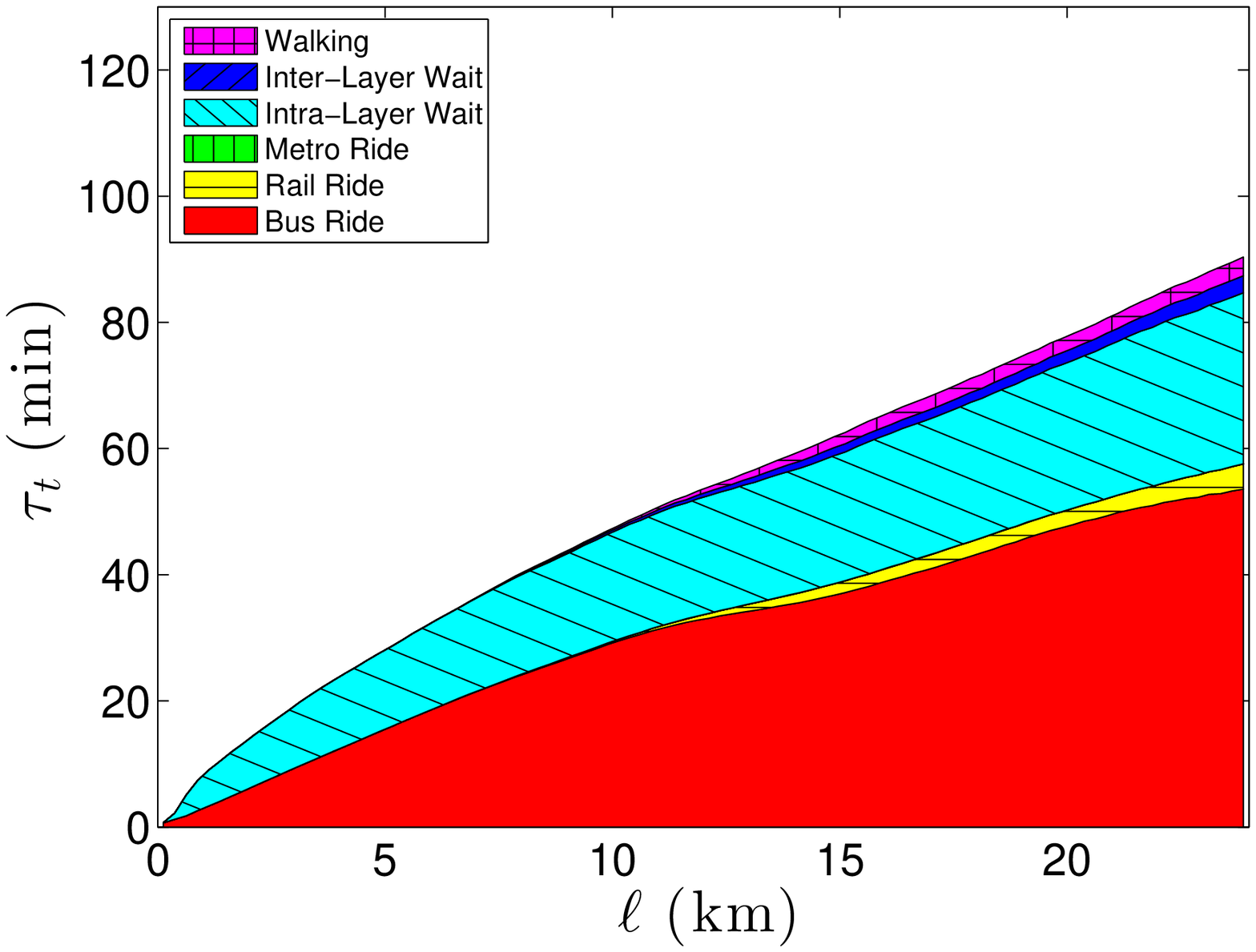} &
\raisebox{2.3cm}{(d)} \includegraphics[angle=0, width=0.45\textwidth]{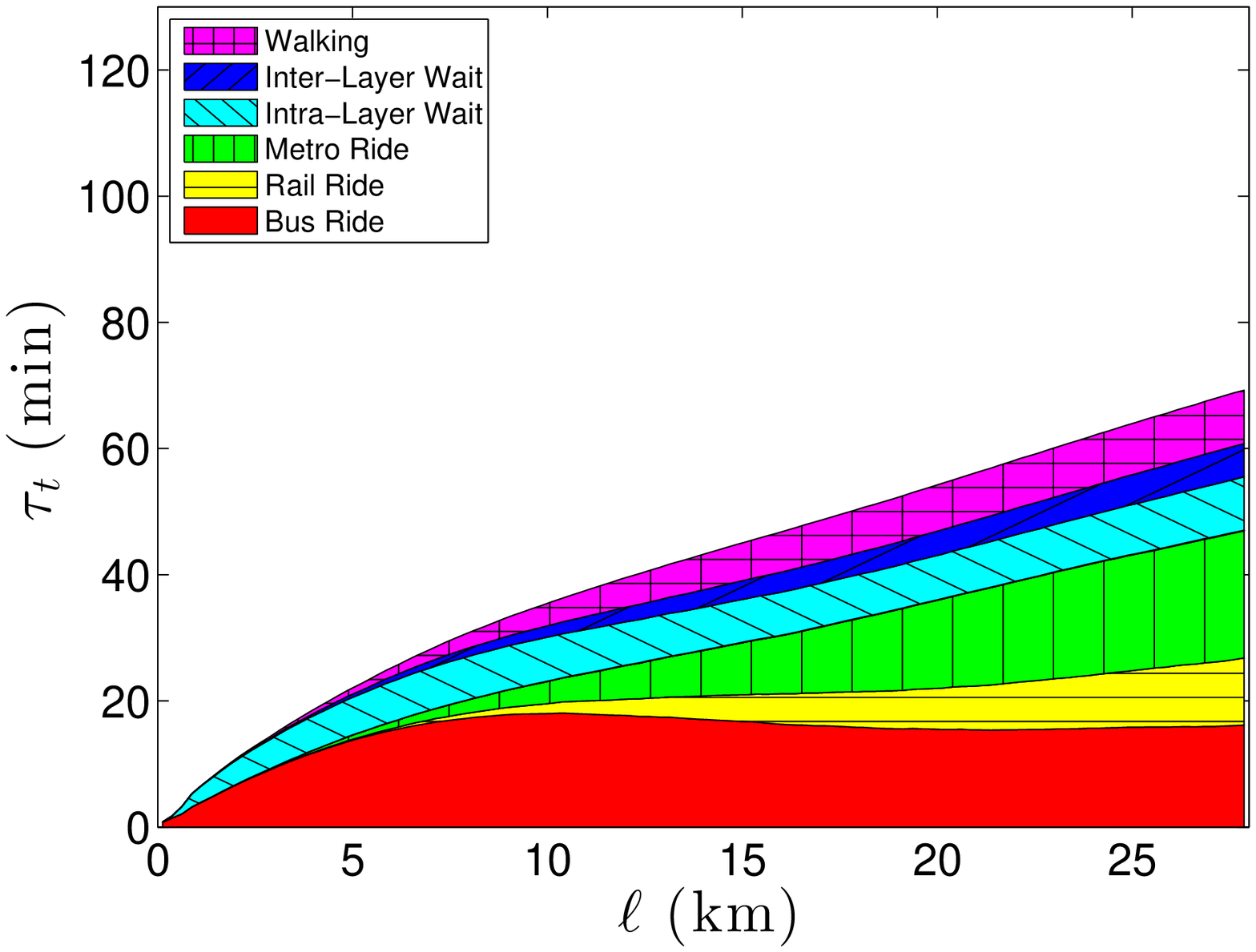} \\
\end{tabular}
\end{center}
\caption{The Anatomy of the transport networks in: (a) Leeds; (b) Liverpool; (c) Sheffield; (d) Inner London}
\end{figure*}

\end{document}